\documentclass[lettersize,journal]{IEEEtran}

\usepackage{amsmath,amssymb,mathtools,derivative,siunitx}
\usepackage{algpseudocode}
\usepackage{dsfont}
\usepackage{tikz,tkz-euclide}
\usepackage{pgfplots}
\usepgfplotslibrary{groupplots}
\usetikzlibrary{calc}
\usepackage[sort&compress]{natbib}
\usepackage{algorithm}
\usepackage{algpseudocode}
\usepackage{xcolor,comment}
\definecolor{rougeG}{rgb}{.76,0,.12}
\definecolor{vertG}{rgb}{.07,.56,.25}
\definecolor{bleuG}{rgb}{.07,.01,.9}
\definecolor{om}{rgb}{.5,.25,.25}
\definecolor{amethyst}{rgb}{0.6, 0.4, 0.8}

\def\CPA{\mbox{CPA}} % CPA
\def\one{\mathds{1}}

\def\zerob{\boldsymbol{0}}
\def\hb{\mathbf{h}}  % normalized magnetic field
\def\mb{\mathbf{m}}  % dipolar moment
\def\nb{\mathbf{n}}  % noise
\def\pb{\mathbf{p}}  % 
\def\rb{\mathbf{r}}  % position vector
\def\sb{\mathbf{s}}  % signal
\def\ub{\mathbf{u}}  % normalized position vector
\def\vb{\mathbf{v}}  % generic vector
\def\xb{\mathbf{x}}  % mesure / observation
\def\Ab{\mathbf{A}}  % coefficient matrix A
\def\Bb{\mathbf{B}}  % magnetic induction field
\def\Cb{\mathbf{C}}  % 
\def\Fb{\mathbf{F}}  % discretized basis
\def\Gb{\mathbf{G}}  % magnetic gradient tensor
\def\Hb{\mathbf{H}}  % magnetic field
\def\Ib{\mathbf{I}}  % identity matrix
\def\Lb{\mathbf{L}}  % Cholesky matrix
\def\Mb{\mathbf{M}}  % Dipolar moment in the matrix modeling H
\def\Pb{\mathbf{P}}  % Orthogonal bases changes
\def\Qb{\mathbf{Q}}  % generic matix Q
\def\Rb{\mathbf{R}}  % rotation matrix
\def\Sb{\mathbf{S}}  % change of base (no necessariliy unitary) for similarity

\def\alphab{\boldsymbol{\alpha}}  % generic vector alpha
\def\betab{\boldsymbol{\beta}}  % generic vector beta
\def\varthetab{\boldsymbol{\vartheta}}  % generic parameter
\def\Pib{\boldsymbol{\Pi}}  % measure operator/matrix
\def\Psib{\boldsymbol{\Psi}} % temporal covariance
\def\Sigmab{\boldsymbol{\Sigma}} % spatial covariance
\def\F{\mathcal{F}}  % signal base
\def\H{\mathcal{H}}  % detection/discrimination hypothesis
\def\L{\mathcal{L}}  % likelihood
\def\N{\mathcal{N}}  % gaussian
\def\V{\mathcal{V}}  % constrained space
\def\P{\mathcal{P}}

\def\tnb{\tilde{\nb}}
\def\tsb{\tilde{\sb}}
\def\txb{\tilde{\xb}}
\def\tAb{\tilde{\Ab}}
\def\tFb{\tilde{\Fb}}
\def\tV{\widetilde{\V}}

\def\hmb{\widehat{\mb}}
\def\hAb{\widehat{\Ab}}
\def\hvarthetab{\widehat{\varthetab}}

\def\Rset{\mathbb{R}}  % reals
\def\SNR{\text{SNR}}  % rsb
\def\dB{\text{dB}}  % dB

\def\argmax{\operatorname*{argmax}}
\def\argmin{\operatorname*{argmin}}
\def\det{\operatorname{det}}
\def\diag{\operatorname{diag}}
\def\div{\operatorname{div}}

\def\Tr{\operatorname{Tr}}
\def\grad{\boldsymbol{\operatorname{grad}}}
\def\vec{\operatorname{vec}}
\def\vect{\operatorname{span}}

\def\Pd{P_{\mathrm{d}}}
\def\Pfa{P_{\mathrm{fa}}}
\def\cdf{F}
\def\ccdf{\bar{F}}

\definecolor{bleu}{HTML}{1f77b4}
\definecolor{orange}{HTML}{ff7f0e}
\definecolor{vert}{HTML}{2ca02c}
\definecolor{rouge}{HTML}{d62728}

\begin{document}
\title{Generalized likelihood ratio test for magnetic anomaly detection: a geometrical approach}
\author{\IEEEauthorblockN{
C. Chenevas-Paule\IEEEauthorrefmark{1}\IEEEauthorrefmark{2}\IEEEauthorrefmark{3}\IEEEauthorrefmark{4},
S. Zozor\IEEEauthorrefmark{1}\IEEEauthorrefmark{4}
L.-L. Rouve\IEEEauthorrefmark{2}\IEEEauthorrefmark{4},
O.~J.~J. Michel\IEEEauthorrefmark{1}\IEEEauthorrefmark{4},
O. Pinaud\IEEEauthorrefmark{2}\IEEEauthorrefmark{4} and
R. Kukla\IEEEauthorrefmark{3}\IEEEauthorrefmark{4}\\}

\IEEEauthorblockA{\IEEEauthorrefmark{1}  Univ.  Grenoble Alpes,  CNRS,  Grenoble
  INP, GIPSA-Lab, 38000 Grenoble, France}
  
\IEEEauthorblockA{\IEEEauthorrefmark{2}  Univ.  Grenoble Alpes,  CNRS,  Grenoble
  INP, G2Elab, 38000 Grenoble, France}
  
\IEEEauthorblockA{\IEEEauthorrefmark{3} Centre  d'Expertise pour la  Maîtrise de
  l'Information et des Signatures, Naval Group, 83100 Ollioules, France}
  
\IEEEauthorblockA{\IEEEauthorrefmark{4}  Naval  Electromagnetism Laboratory,  21
  avenue des Martyrs, 38000 Grenoble, France}
%\thanks{Manuscript created October,  2020; This work was developed  by the IEEE
%Publication Technology  Department. This work  is distributed under  the \LaTeX
%\  Project  Public License  (LPPL)  (  http://www.latex-project.org/ )  version
%1.3.  A  copy  of the  LPPL,  version  1.3,  is  included in  the  base  \LaTeX
%\  documentation  of all  distributions  of  \LaTeX  \ released  2003/12/01  or
%later. The opinions expressed here are entirely that of the author. No warranty
%is expressed or implied. User assumes all risk.}
}

\markboth{Journal of \LaTeX\ Class Files,~Vol.~18, No.~9, September~2025}%
{How to Use the IEEEtran \LaTeX \ Templates}

\maketitle

\begin{abstract}
State-of-the-art  approaches   to  magnetic   anomaly  detection  rely   on  the
generalized likelihood  ratio test  (GLRT).  These approaches  are based  on the
formulation of a parametric  model of the source to be  detected, expressed in a
suitable functional  basis. One of  the primary objectives  of this study  is to
demonstrate  that,  for  a  given   measurement  configuration,  the  signal  is
constrained to evolve within a restricted subset of the space generated by these
functional bases. The parametric representation of the signal is identified as a
semi-algebraic space which, for the dipole model used in this article, turns out
to be a cone outside of which the estimated signal does not satisfy the physical
equations.  Thus,  a second objective is  to exploit this property  to constrain
the signal  parameters in  the GLRT  to belong to  the semi-algebraic  space, in
order to  improve detection  performance. The performance  gain of  the proposed
algorithm  is  compared  to  the   one  of  conventional  approaches;  numerical
simulations   show   that   the   proposed   approach   not   only   outperforms
state-of-the-art methods  but can  even provide  results close  to those  of the
clear-seeing (optimal) receiver.
\end{abstract}

\begin{IEEEkeywords}
Magnetic   anomaly   detection,   Dipolar  models,   constrained   optimization,
Generalized likelihood ratio test, semi-algebraic space, detection performances.
\end{IEEEkeywords}

% ------------------------------- Introduction ------------------------------- %

\section{Introduction}
\IEEEPARstart{A}  magnetic object  immersed in  a  field reacts  by creating  an
induced     field     that     disturbs     the     field     that     generated
it~\cite{Straton2007Electromagnetic}.    In   addition,  such   objects   (e.g.,
ferromagnetic) may  have their own  permanent magnetic  field. The sum  of these
fields therefore appears  to the observer as a magnetic  anomaly compared to the
reference  situation in  the absence  of the  object.  Detecting  such anomalies
therefore leads to the detection of magnetic objects, and has motivated numerous
studies referred  to as MAD (for  Magnetic Anomaly Detection). MAD  has numerous
applications,  such  as the  search  for  underwater  pipes or  cables,  wrecks,
submarines,              etc~\cite{Loane1976speed,              Blanpain1979PhD,
  Zhao2021MADReview}. Historically,  a magnetic sensor  is moved within  a scene
containing  a supposedly  fixed  target,  seeking to  detect  any local  spatial
variations  in  the  Earth's  magnetic  field that  might  be  produced  by  the
target~\cite{MADProgram1946,       Loane1976speed,      Fitterman1987MADSurveys,
  Wynn1999DetectionStaticDipole}.  Of course, a moving target and a fixed sensor
lead   to  a   similar  detection   problem~\cite{Wynn1999DetectionStaticDipole,
  Zhao2021MADReview}.

Detecting the presence  of an object using magnetic signal  records gave rise to
many   technical  developments~\cite{MADProgram1946,   Zhao2021MADReview}.   For
instance noise-based  approaches consist in  detecting changes in the  nature of
the    noise    statistics,    thus     revealing    the    presence    of    an
anomaly~\cite{Qiao2023adaptive,  Sheinker2008magnetic}.    More  recently,  many
attempts were also deployed  based on learning approaches~\cite{Liu2019magnetic,
  Fan2020adaptive,       Hu2020magnetic,      Xu2020deepmad,       Wang2022deep,
  Chen2024magnetic}.

This paper  focuses on the  historical approaches, quite  recently re-discovered
and    generalized,    based   on    physical    modeling    of   the    anomaly
signal~\cite{MADProgram1946,           Loane1976speed,          Blanpain1979PhD,
  Ginzburg2002processing,      Sheinker2009magnetic,     Pepe2015generalization,
  Fan2020gradient,  Qin2020magnetic,  Chenevas2024analytical,  Yan2024effective,
  Chenevas2026MAD}.  Under certain assumptions on  the trajectory of the sensor,
the  magnetic anomaly  signal can  be expanded  into a  class of  functions; the
expansion weights depend solely on the  unknown source, while the function basis
depends only on the geometric configuration.   Thus, in the presence of additive
noise  (sensor  noise, environmental  noise,  etc.),  the detection  problem  is
reformulated as a classical problem of detecting a parametric signal (parameters
involved  both in  the model  of expansion  functions and  weights) embedded  in
noise~\cite{Kay1998detection,  vanTrees2013DetectionEstimation}. In  general, no
restriction  is imposed  to  the  parameters of  the  signal  when detection  is
performed, and, for  instance, a generalized (log-)likelihood  ratio test (GLRT)
is performed.

In this paper, model-based MAD is reexamined; specifically, the structure of the
space  covered by  all physically  possible  signals is  studied.  An  important
contribution  consists in  showing  that for  given  situations, the  observable
anomaly signals live on a semi-algebraic  sub-space, but not in the whole vector
space spanned by the functions basis.  This  is accounted for in order to derive
an improved GLRT.

The paper  is organized  as follows.  In  section~\ref{Sec:Problem}, geometrical
configuration,  source  modeling,  recorded   anomaly  signals  under  different
configurations  (scalar   or  3D   sensors),  and   noise  are   introduced.  In
section~\ref{Sec:DetectionProblem}, a brief reminder on GLRT and its performance
is  provided, when  no  constraint is  imposed on  the  signal parameters.  This
strategy  is  reexamined   in  section~\ref{Sec:PhysicallyConstrainedGLRT}  when
constraints on  the parameters imposed  by the  physics are taken  into account.
Finally,  the new  original strategy  incorporating the  aforementioned physical
constraints  is  assessed   in  section~\ref{Sec:Simulation}  through  numerical
simulations, and it is shown that it significantly outperforms the unconstrained
GLRT   approach.    Some  discussions   and   perspectives   are  developed   in
section~\ref{Sec:Conclusion}.

% ---------------------------- Problem Statement ----------------------------- %

\section{Problem statement}
\label{Sec:Problem}

% ------------------------ Geometry

\subsection{Geometry of the problem}
\label{SubSec:Geometry}

\begin{figure}[ht]
\begin{center}
\begin{tikzpicture}[scale=1.5]

%% some definitions
\pgfmathsetmacro{\El}{25} % elevation
\pgfmathsetmacro{\Az}{-135}

% Projection pour le trace d'un point (x,y,x) compte tenu de l'angle de vision
\pgfmathdeclarefunction{projx}{2}{\pgfmathparse{#1*cos(\Az)-#2*sin(\Az)}}
\pgfmathdeclarefunction{projy}{3}{\pgfmathparse{#1*sin(\Az)*sin(\El)+#2*cos(\Az)*sin(\El)+#3*cos(\El)}}

\pgfmathsetmacro\alphaRef{35} % angle alpha
\pgfmathsetmacro\cR{cos(\alphaRef)} % cos alpha
\pgfmathsetmacro\sR{sin(\alphaRef)} % sin alpha
\pgfmathsetmacro\betaCPA{30} % angle beta
\pgfmathsetmacro\cB{cos(\betaCPA)} % cos alpha
\pgfmathsetmacro\sB{sin(\betaCPA)} % sin alpha
%
% Distance CPA
\pgfmathsetmacro\DCPA{1.7} % D CPA
%
% Point CPA
\pgfmathsetmacro\yCPA{-\DCPA*\sB}
\pgfmathsetmacro\zCPA{\DCPA*\cB}
%
% Point P sur la trajectoire
\pgfmathsetmacro\xP{-2.6}
\pgfmathsetmacro\yP{\yCPA}
\pgfmathsetmacro\zP{\zCPA}

% ==========================================

% Vue perspective
% ===============
%
%
% Trace des axes
% --------------
% Reference (x', son ortogonal dans plan, z(
%
% x' -> x par rotation de - alpha / z
% z -> z' par rotation de beta / x'
% y' par la rotation de (0,1,0) de beta / x'
%
% x' parallele a la trajectoire
\draw[>=stealth, dashed,  ->]
({projx(-.25,0)},{projy(-.25,0,0)})--
({projx(2.8,0)},{projy(2.8,0,0)})
node[left,scale=.9]{$x'$};
%
% x choisi
\draw[>=stealth, ->]
({projx(-.25*\cR,.25*\sR)},{projy(-.25*\cR,.25*\sR,0)})--
({projx(2.3*\cR,-2.3*\sR)},{projy(2.3*\cR,-2.3*\sR,0)})
node[left,scale=.9]{$x$};
%
% y' orthogonal a la trajectoire et z'
\draw[>=stealth, dashed, ->]
({projx(0,-.25*\cB)},{projy(0,-.25*\cB,.25*\sB})--
({projx(0,1.5*\cB)},{projy(0,1.5*\cB,-1.5*\sB)})
node[right,scale=.9]{$y'$};
%
% y choisi
\draw[>=stealth, ->]
({projx(-.25*\sR,-.25*\cR)},{projy(-.25*\sR,-.25*\cR,0)})--
({projx(2.4*\sR,2.4*\cR)},{projy(2.4*\sR,2.4*\cR,0)})
node[below,scale=.9]{$y$};
%
% z, vertical
\draw[>=stealth, ->]
(0,{projy(0,0,-.1)})--
(0,{projy(0,0,2.4)})
node[above,scale=.9]{$z$};
%
% z', O-CPA
\draw[>=stealth, dashed, ->]
({projx(0,.25*\sB)},{projy(0,.25*\sB,-.25*\cB)})--
({projx(0,-2.3*\sB)},{projy(0,-2.3*\sB,2.3*\cB)})
node[above,scale=.9]{$z'$};
\node[scale=.8] at (0,0) {$\bullet$};
\node[below,scale=.8] at (-.1,-.1) {$O$};
%
%
% Trace de la trajectoire
% ------------------------
%
\draw[thick, dashdotted, ->, >=stealth]
({projx(-3,\yCPA)},{projy(-3,\yCPA,\zCPA)})--
({projx(2.25,\yCPA)},{projy(2.25,\yCPA,\zCPA)})
node[above, scale = .9]{$\gamma$};
%
%
% Point CPA
% ----------
%
\coordinate (S) at (0,0); % Source
\coordinate (CPA) at ({projx(0,\yCPA)},{projy(0,\yCPA,\zCPA)}); % CPA
\coordinate (PCPA) at ({projx(0,\yCPA)},{projy(0,\yCPA,0)}); % projection CPA sur le plan horizontal
\node[scale=.8] at (CPA) {$\bullet$}; % positionnement du CPA
\node[above left,scale=.8] at (CPA) {CPA ($t=t_0$)\:};
\draw[dotted] (CPA)--(PCPA)--(S) % ligne de projection CPA-plan - source
pic [draw=black, solid, scale=.4] {right angle = CPA--PCPA--S};% angle droit marqué
\node[below,scale=.8] at ({projx(0,.6*\yCPA)},{projy(0,.5*\yCPA,.5*\zCPA)}) {$D$}; % placement de D
%
%
% point P
%
\coordinate (P) at ({projx(\xP,\yP)},{projy(\xP,\yP,\zP)}); % P
\node[scale=.8] at (P) {$\bullet$}; % positionnement de P
\node[above left,scale=.8] at (P) {$P$};
\draw[dotted, ->, line width=0.3mm, >=latex] (0,0)--(P); % ligne O-P
\node[right,scale=.8] at ({0.5*projx(\xP,\yP)},{0.5*projy(\xP,\yP,\zP)}) {$\rb$}; % placement de rb
%
%
% Angle beta
\coordinate (A) at ({projx(0,0)},{projy(0,0,\zCPA)});
\coordinate (C) at ({projx(0,\yCPA)},{projy(0,\yCPA,\zCPA)});
\tkzMarkAngle[size=.9*\zCPA, ->, >=stealth,mark=none](A,S,C);
\tkzLabelAngle[pos=1.25*\zCPA,scale=.8](A,S,C){$\beta$}
%
%
% Angle alpha
\coordinate (A) at ({projx(\cR,-\sR)},{projy(\cR,-\sR,0)});
\coordinate (C) at ({projx(1,0)},{projy(1,0,0)});
\tkzMarkAngle[size=\zCPA, ->, >=stealth, mark=none](A,S,C)
\tkzLabelAngle[pos=1.4*\zCPA,scale=.8](A,S,C){$\alpha$}
%
%\node at (A) {$\bullet$};
%\node at (C) {$\bullet$};
\end{tikzpicture}
\end{center}
\caption{Geometry and notations of the problem. The path travelled by the sensor
  (trajectory  $\gamma$) is  represented  by the  dash-dotted  line. The  sensor
  position is marked both by its position $P$ on $\gamma$ and by the vector $\rb
  = \boldsymbol{OP}$.}
\label{Fig:Geometry}
\end{figure}
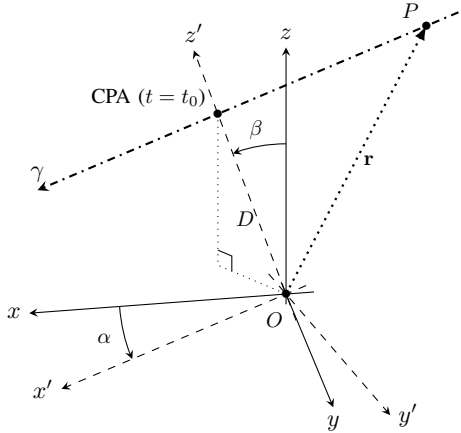

The   case  of   a  moving   sensor   and  a   fixed  target   is  depicted   in
Fig.~\ref{Fig:Geometry}.  It is  assumed that  the sensor's  trajectory $\gamma$
(set of  successive sensor positions $P(t)$  on the figure) is  a straight line,
travelled at constant speed $V$ and altitude.  $V$ is assumed to be much greater
than the  possible velocity of  the target, so this  latter is considered  to be
stationary.  We therefore adopt a  direct orthogonal reference frame centered on
the source $O$, in which $Oz$ is the vertical axis.  Then, we arbitrarily define
the $x$ and $y$ axis  in the horizontal plane, such that $O x  y z$ is the world
coordinate system  which is known.   The point from  the trajectory that  is the
closest to the  source is called CPA  for ``closest point of  approach''; let be
$t_0$  the time  when the  sensor is  at the  CPA, and  let $D$  be the  minimal
distance between $O$ and $P$, i.e., when $P(t) = \CPA = P(t_0)$.  Let us finally
introduce the $z'-$ axis to be the  oriented axis containing $O$ and the $\CPA$;
let  $x'-$axis be  the horizontal  axis  collinear to  the trajectory  $\gamma$;
$y'-$axis is  defined so as  to form an oriented  orthogonal $x'y'z'-$coordinate
system.   We finally  denote  by $\alpha$  the angle  between  the $x-$axis  and
$x'-$axis and by $\beta$ the angle the $z-$axis and the $z'-$axis.

In the dual case, where the source moves whereas the sensor is fixed, the source
is assumed moving at a constant speed and constant altitude along $Ox'$, so that
we can  still virtually  center the  coordinate system to  the source.  The main
difference  is that  in the  first  case $\alpha$  is known,  and therefore  the
orientation of the coordinate system can be chosen so that $\alpha = 0$, whereas
it is unknown in the second situation.
  
In the $O x'y'z'-$ reference frame,  $\rb(t)$ is expressed as $\begin{bmatrix} V
  \,  (t-t_0) &  0 &  D \end{bmatrix}^\top$;  We denote  by $\Rb_u(\theta)$  the
rotation matrix of angle $\theta$ around  the $u-$axis (in some given coordinate
system),  and let  $\rb  \equiv  \rb_\gamma$ when  the  sensor  moves along  the
trajectory, then
\begin{equation}
\rb_\gamma : t 
\mapsto
\Rb_z(\alpha) \, \Rb_x(\beta)  \begin{bmatrix}
V \, (t-t_0) \\
0 \\
D
\end{bmatrix}
\label{Eq:Trajectory}
\end{equation}
with
\begin{equation*}
\alpha \in \left[-\pi  \, , \, \pi  \right], \quad \beta \in \left[-\pi  \, , \,
  \pi \right].
\end{equation*}
In the case of a sensor on board an aircraft, the latter necessarily passes above
the target, so $\beta$ can  be limited  to belong to $\left[  -\frac\pi2 \,  , \,
  \frac\pi2 \right]$, and $\alpha$ can be arbitrarily set to $0$.

% ------------------------ Source modeling

\subsection{Source modeling}
\label{SubSec:SourceModeling}

% Maxwell-Ampere Curl H = J + partial D / partial t
% Maxwell-Thomson Div B = 0
According to  the Maxwell  equations in  a static framework,  in the  absence of
current  and charge,  the  magnetic field\footnote{Actually  the magnetic  field
$\Hb$  is  expressed in  Amp\`eres  per  meter,  while  this is  the  ``magnetic
induction  field'' $\Bb$,  expressed in  Tesla, which  satisfies $\div  \Bb =0$.
However, in  homogeneous isotropic  medium the  latter field  is related  to the
former by $\Bb = \mu \Hb$ with $\mu$ the magnetic permeability, not depending on
the spatial coordinates ($\mu = \mu_0 = 4 \pi 10^{-7}$ ).} $\Hb$ can be shown to
be expressed  as a function of  a scalar potential $\psi$  as $\Hb = -  \grad \,
\psi$.  As  $\div \Hb  = 0$, the  scalar potential satisfies  $\Delta \psi  = 0$
(Laplace    equation)    where    $\Delta$     stands    for    the    Laplacian
operator~\cite{Straton2007Electromagnetic}.  The solution is well known, usually
expressed  in   spherical  coordinates  $\rb(r,\theta,\varphi)$.    Outside  the
Brillouin  sphere (the  smallest sphere  containing  the source),  it allows  to
express the magnetic field generated by the  source (the anomaly) in the form of
the following decreasing multipolar  expansion\footnote{All the divergent source
terms (given  for $n <  0$) have disappeared  from the expression;  the monopole
term  (given   for  $n=0$)   cancels  as   well,  since   it  has   no  physical
reality~\cite{Straton2007Electromagnetic}.}~\cite[Sec.~3.12]{Straton2007Electromagnetic}:
\begin{equation*}
\begin{split}
\Hb(\rb) = - \nabla \: \sum_{n=1}^{+\infty} & \dfrac1{r^{n+1}} \sum_{m=0}^n
\Big( a_{n,m} \cos(m \varphi) \\
& +  b_{n,m} \sin(m \varphi) \Big) \P_n^m(\cos \theta)
\end{split}
\raisetag{2\normalbaselineskip}
\end{equation*}
where $\P_n^m$ stand for the Legendre function  of degree $n$ and order $m$; the
harmonic coefficients $a_{n,m}, \, b_{n,m}$ depend on the magnetic source and on
the coordinate  system in which $\rb$  is expressed.  The first  term ($n=1$) in
the expansion corresponds to a dipole,  the second term ($n=2$) to a quadrupole,
etc~\cite{Straton2007Electromagnetic}.  Note  that $b_{n,0}$ is  irrelevant and,
without loss of generality, can be set to $0$.
% \begin{figure}[ht]       \input{brillouin}       \caption{Two-dimensional
% representation of the Brillouin sphere and the validity zone of the decreasing
% harmonic decomposition. The  greyed-out area represents the  source, while the
% dotted  circle is  the  Brillouin sphere.  The  decreasing spherical  harmonic
% expansion is valid outside this sphere.}  \end{figure}

At a  sufficiently large distance from  the source, the magnetic  anomaly can be
approximated  by  the  first  term  of  the  multipolar  decomposition,  as  all
$2^n-$polar terms decay as $\frac1{r^{n+2}}$. The dipolar approximation reads:
\begin{equation}
\label{Eq:Dipole}
\Hb(\rb) = \dfrac{3 \left( {\mb}^\top \ub_r \right) \ub_r - \mb}{r^3}
\end{equation}
%
% \mn en A m^2
where   $\mb   =   \begin{bmatrix}   m_x  &   m_y   &   m_z   \end{bmatrix}^\top
=  \begin{bmatrix}  -a_{1,1}  &  -b_{1,1} &  a_{1,0}  \end{bmatrix}^\top$  is  the
so-called     dipolar      magnetic     moment~\cite{Straton2007Electromagnetic,
  Wikwo1984Multipole}.  In  the equation  above, $r  = \| \rb  \|$ and  $\ub_r =
\dfrac{\rb}{r}$.   Throughout the  rest of  this study,  this realistic  dipolar
approximation is adopted.

\begin{figure}[ht]
\begin{center}
\begin{tikzpicture}[decoration={markings,mark=at position 1.5cm with {\arrow[blue]{stealth};}}]
	\clip circle(3);
	\foreach \K in{1.5,2.9,6,12}{
	\draw [postaction=decorate,thin, color=blue, 
	domain=pi/2:-pi/2, samples=50]plot (xy polar cs:angle=\x r, radius={\K*cos(\x r)*cos(\x r)});
	\draw [postaction=decorate,thin, color=blue, domain=pi/2:3*pi/2,samples=50]plot (xy polar cs:angle=\x r, radius={\K*cos(\x r)*cos(\x r)});	
	}	
	\draw[postaction=decorate,thin,color=blue](0,0)--(0,3);
	\draw[postaction=decorate,thin,color=blue](0,-3)--(0,3);
	\draw[dashed,fill=white]circle(0.5);
	\draw[<-] (0,5pt)--++(0,-10pt)node[midway,right]{\footnotesize $\mb$};
\end{tikzpicture}
\end{center}
\caption{Field lines around the magnetic dipole}
\end{figure}
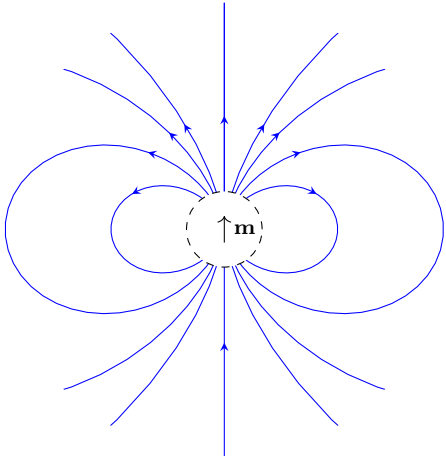

We will  see later  that the  gradient tensor (Jacobian  matrix) of  the dipolar
field $\Gb \equiv  \nabla \; \Hb^\top$ can  be used in detection  schemes; it is
expressed as follows:
\begin{equation}
\label{Eq:Gradient}
\Gb(\rb) = \dfrac3{r^4}  \Big( \ub_r \, \mb^\top + \mb  \, \ub_r^\top + \mb^\top
\ub_r \, \left( \Ib_3 - 5 \,\ub_r \, \ub_r^\top \right) \Big)
\end{equation}
where  $\Ib_k$  stands  for  the  $k   \times  k$  identity  matrix.   $\Gb$  is
symmetrical\footnote{This is  a consequence of the  Maxwell-Amp\`ere equation.}.
Furthermore, as $\Tr \Gb = \div \Hb  = 0$, the gradient tensor is symmetric with
zero  trace, and  depends on  $5$ coefficients  only ($2$  on the  diagonal, $3$
outside), parameterized by the dipolar moment $\mb$.

% ------------------------ Signal modeling

\subsection{Signal modeling}
\label{SubSec:SignalModeling}

In this  study, it is  assumed that the Earth's  magnetic field is  known (e.g.,
measured using  a sensor of reference  sufficiently far from the  anomaly). Thus
the Earth's field can be removed from the sensor measurements.

In order to  derive the ideal noise-free  signal expected to be  recorded by the
sensor along  the trajectory  $\gamma$, an operator  $\Pib$ is  introduced. Note
that $\Pib$ encompasses all operations that allow to relate the recorded anomaly
signal  to the  magnetic  field  anomaly in  the  world $xyz-$coordinate  system
(examples will  be given in  sub-subsection~\ref{SubSubSec:VectorMeasurement} to
\ref{SubSubSec:InvariantsMeasurement}). Therefore:
\begin{equation*}
\sb = \Pib \circ \Hb\left(\rb_\gamma\right)
\end{equation*}
where  $\circ$ denotes  the operational  relation (e.g.,  matrix multiplication,
square operation, etc.).

In  the  $x'y'z'-$coordinate  system,  the unitless  variable  and  the  unitary
``radial'' vector along the trajectory are introduced:
\begin{equation}
\label{Eq:uPrime}
u    =     \dfrac{V    \,    (t     -    t_0)}{D},    \quad     \ub'_\gamma    =
\frac1{\sqrt{1+u^2}} \begin{bmatrix} u\\0\\1 \end{bmatrix}
\end{equation}
Thus,  along the trajectory,  from Eq.~\eqref{Eq:Trajectory},
\begin{equation}
\label{Eq:rTrajectory}
\rb_\gamma(t) = r_\gamma \, \Rb_z(\alpha) \, \Rb_x(\beta) \, \ub'_\gamma , \quad
r_\gamma = D \,\sqrt{1+u^2}
\end{equation}
Then, the  expression of  the dipolar  moment in the  same coordinate  system as
$\ub'_\gamma$ is
\begin{equation}
\mb'  =  \Rb_x(\beta)^\top  \,   \Rb_z(\alpha)^\top  \,  \mb  =  \begin{bmatrix}
  m_x'\\[1mm] m_y'\\[1mm] m_z'\end{bmatrix}
\end{equation}
and the magnetic field along the trajectory Eq.~\eqref{Eq:Dipole} reads:
\begin{equation}
\label{Eq:AnomalyTrajectory}
\Hb(\rb_\gamma(t))   =   \frac{\Rb_z(\alpha)   \,  \Rb_x(\beta)   \,   \Big(   3
  \left(\mb'^\top  \ub'_\gamma  \right)  \,  \ub'_\gamma -  \mb'  \Big)}{D^3  \,
  (1+u^2)^{\frac32}}
\end{equation}

Similarly,  $\Gb(\rb_\gamma(t))$  is  easily  obtained from  the  expression  in
equation~\eqref{Eq:Gradient}   by  left-multiplication   by  $\Rb_z(\alpha)   \,
\Rb_x(\beta)$,  right-multiplication by  $\Rb_x(\beta)^\top \Rb_z(\alpha)^\top$,
and by replacing $\ub_r$ and $\mb$ by $\ub'_\gamma$ and $\mb'$ respectively.
 
The magnetic field related to the anomaly (or its gradient) along the trajectory
is  now parameterized  by  $\mb', \alpha,  \beta,  t_0$ and  $D$,  that must  be
determined.     In    what    follows,    (following    many    other    authors
\cite{Blanpain1979PhD,         Chenevasdetection,        Chenevas2024analytical,
  Chenevas2026MAD, Chenevas2025physics, Fan2020gradient, Ginzburg2002processing,
  Pepe2015generalization,         Qin2020magnetic,         Sheinker2009magnetic,
  Yan2024effective, Zhao2021MADReview}), both  $D$ and $t_0$ will  be assumed to
be known\footnote{In practice,  working with a sliding windows, we
will assume that $t_0$  is at the center of the windows, which  will be the true
value when  the sensor achieve  the CPA. The estimation  of $D$ remains  and its
estimation is left as a perspective.\label{Foot:operational}}.  As a consequence
the reduced variable  $u$ is known, and  both the field and  its gradient tensor
decompose onto  a set of  functions of $u$.  This  will be largely  explored and
used in the sections to come.  \

% -------------- Vector measurement

\subsubsection{$d-$axes measurement of the anomaly}
\label{SubSubSec:VectorMeasurement}

A $d-$axes  magnetometer measures  the projection  of the  field onto  these $d$
axes; thus, denoting by $\Pib$ the $\Rset^{3 \times d}$ matrix whose columns are
vectors oriented  along the projection  axes (expressed in  the $xyz-$coordinate
system),  the  signal   recorded  by  the  sensor  reads   $\sb(t)  =  \Pib^\top
\Hb(\rb_\gamma(t))$.       Then,      after       some      simple      algebra,
Eq.~\eqref{Eq:AnomalyTrajectory} can be re-expressed as
\begin{equation}
\label{Eq:BasisAnomaly}
\Hb(u)  \equiv \Hb(\rb_\gamma(t))  = \Rb_z(\alpha)  \, \Rb_x(\beta)  \, \Mb'  \:
\frac1{(1+u^2)^{\frac52}} \begin{bmatrix} 1\\u\\u^2\end{bmatrix}
\end{equation}
where
\begin{subequations}
\begin{align}
\label{Eq:MMatrixdAxis}
\Mb' & = \frac1{D^3}\begin{bmatrix}
 - m'_x   & 3 \, m'_z & 2 \, m'_x \\[2mm]
 - m'_y   &     0     &   - m'_y  \\[2mm]
2 \, m'_z & 3 \, m'_x &   - m'_z
\end{bmatrix}\\[2mm]
\label{Eq:MKroneckerdAxis}
& = \Cb \, \big( \Ib_3 \otimes \mb' \big)
\end{align}
\end{subequations}
with $\otimes$ the Kronecker product and
\begin{equation}
\Cb = \frac1{D^3} \, \begin{bmatrix*}[r]
-1 &  0 & 0 & 0 & 0 & 3 & 2 &  0 &  0\\[2mm]
 0 & -1 & 0 & 0 & 0 & 0 & 0 & -1 &  0\\[2mm]
 0 &  0 & 2 & 3 & 0 & 0 & 0 &  0 & -1
\end{bmatrix*}
\end{equation}
Thus the measured signal reads
\begin{equation}
\label{Eq:ProjectedAnomaly}
\sb(u) = \Ab \, \frac{1}{(1+u^2)^\frac52} \: \begin{bmatrix}
1\\[1mm]
u\\[1mm]
u^2
\end{bmatrix}
\end{equation}
with
\begin{subequations}
\label{Eq:CoeffProjectedAnomaly}
\begin{align}
\label{Eq:CoeffProjectedAnomalyPi}
\Ab & = \Pib^\top \, \Rb_z(\alpha) \, \Rb_x(\beta) \, \Mb'\\[2mm]
\label{Eq:CoeffProjectedAnomalyPip}
& = {\Pib'}^\top \, \Mb'
\end{align}
\end{subequations}
where
\begin{equation}
\Pib' = \Rb_x(\beta)^\top \, \Rb_z(\alpha)^\top \, \Pib
\end{equation}
$\Pib'$   is    the   projective    measurement   matrix   expressed    in   the
$x'y'z'-$coordinate system.

For example,  usual triaxial sensors  are composed  by three orthogonal  axes so
that, if  well calibrated (ideal  case with same gain  on each axis),  $\Pib$ is
(proportional to) a $3 \times 3$ rotation matrix.  As the focus is on the moving
sensor/fixed target scenario,  remind that $\alpha$ can be set  to zero; in such
an  ideal case  pre-multiplying the  measure by  a rotation  matrix in  order to
virtually align its axes with the  reference frame leads to $\Pib \equiv \Ib_3$.
Note however that in  the case of airborne sensors, the  measure is sensitive to
roll (rotation about  the longitudinal), pitch (rotation about  the lateral) and
yaw  (rotation  about  the  vertical)  of the  aircraft,  making  this  approach
difficult  to  implement (in  the  fixed  sensor/moving target  situation,  such
effects do not exist but $\alpha$ is unknown).

Generally  speaking, the  recorded signals  live in  the $\Rset^d-$vector  space
$\vect \F$ where the basis $\vect \F$ is defined as:
\begin{equation}
\label{Eq:BasisVector}
\F = \left\{ u \mapsto \dfrac{u^i}{(1+u^2)^{\frac52}} \right\}_{i=0}^2
\end{equation}
Let us  emphasize that $\vect \F$  is perfectly determined, whereas  the unknown
are entirely embedded in $\Ab \equiv \Ab(\mb',\alpha,\beta)$.

\

% -------------- Scalar measurement

\subsubsection{Scalar measurement of the anomaly}
\label{SubSubSec:ScalarMeasurement}
Scalar magnetometer based measurements correspond to  $\Pib \circ \Hb = \| \Hb +
\Hb_0 \|  - \| \Hb_0  \|$, where  $\Hb_0$ is the  Earth's field (remember  it is
assumed to  be known, see beginning  of subsection~\ref{SubSec:SignalModeling}).
Assuming that the  anomaly is small compared to the  Earth's field, $\|\Hb\| \ll
\|\Hb_0\|$, a first order Taylor expansion of the measured total field $\| \Hb +
\Hb_0 \|$ gives
\begin{eqnarray*}
\|  \Hb   +  \Hb_0   \|  &   =  &  \|\Hb_0\|   \,  \sqrt{1   +  2   \,  \Hb^\top
  \frac{\Hb_0}{\|\Hb_0\|} + \frac{\|\Hb\|^2}{\|\Hb_0\|^2}}\\
& =  & \|  \Hb_0 \|  + \Hb^\top \hb_0  + o  \left( \dfrac{\|  \Hb \|}{\|\Hb_0\|}
\right)
\end{eqnarray*}
where the normalized earth's field is introduced:
\begin{equation}
\hb_0 = \frac{\Hb_0}{\| \Hb_0 \|}
\end{equation}
The operator  $\Pib$ can be approximated  (linearized) by an inner  product with
$\hb_0$;      the      measured      signal     is      still      given      by
Eqs.~\eqref{Eq:ProjectedAnomaly}~\eqref{Eq:CoeffProjectedAnomaly} with
\begin{equation}
\Pib  = \hb_0
\end{equation}
Let  us remark  that  such  scalar measurement  remain  unaffected by  rotation,
including yaw, pitch  and roll experienced in airborne contexts,  which may make
them valuable in practice.

\

% -------------- Square modulus measurement

\subsubsection{Square modulus of the anomaly}
\label{SubSubSec:SquareModulusMeasurement}
Building  on the  preceding remark,  another simple  invariant (with  respect to
rotations) is  the square modulus  of the anomaly  (again we assume  that before
taking the  modulus the  Earth's field $\Hb_0$  is removed).   The corresponding
operator $\displaystyle \Pib$  is the squared norm $\| \cdot  \|^2$.  The use of
the square modulus has already been studied in~\cite{Sheinker2009magnetic}; from
expression~\eqref{Eq:AnomalyTrajectory}, we obtain:
\begin{equation*}
\left\|  \Hb \right\|^2  =  \frac{3 \left(  {\mb'}^\top  \ub_\gamma' \right)^2  +
  {\mb'}^\top \mb'}{D^6 \left( 1 + u^2 \right)^3}
\end{equation*}
This leads to the expression of the measured signal:
\begin{equation}
\sb(u) =  \Ab \, \frac{1}{(1+u^2)^4} \: \begin{bmatrix}
1\\[1mm]
u\\[1mm]
u^2
\end{bmatrix}
\end{equation}
with
\begin{equation}
\label{Eq:CoeffSquareModulus}
\Ab  = \frac1{D^6} \begin{bmatrix}
{m'_x}^2 + {m'_y}^2 + 4 \, {m'_z}^2 \\[2mm]
6 \, m'_x \, m_z'\\[2mm]
4 \, {m'_x}^2 + {m'_y}^2 + {m'_z}^2
\end{bmatrix}^\top
\end{equation}
Here again, the measured signal decomposes on a basis that does not depend on the
unknown variables to be estimated; this basis is
\begin{equation*}
\label{Eq:BasisSquareModulus}
\F = \left\{ u \mapsto \dfrac{u^i}{(1+u^2)^4} \right\}_{i=0}^2
\end{equation*}
The unknown  dipolar moment  $\mb'$ is  (again) hidden  in the  coefficient $\Ab
\equiv \Ab(\mb')$.  Note however that the magnitude of the signal scales here as
$\left(  \frac{\| \mb  \|}{D^3} \right)^2$  instead of  $\frac{\| \mb  \|}{D^3}$
precedingly:  the advantage  brought by  invariance to  $\alpha, \beta$  is thus
balanced by  an important increase  of the  measure sensitivity to  the distance
$D$.
  
\

% -------------- Invariants of the magnetic gradient

\subsubsection{Principal invariants of the magnetic gradient}
\label{SubSubSec:InvariantsMeasurement}

The  interest for  using invariants  seems obvious  from the  discussion of  the
preceding subsection. A more constructive  approach is sketched in this section,
that leads to introduce the magnetic tensor gradient $\Gb$. This latter is often
used  in   the  MAD   framework~\cite{Qin2020magnetic,Yan2024effective}.   Since
$\Hb_0$ is assumed to be constant, its gradient tensor is zero, so its knowledge
is unnecessary.

Similarity invariant is a function $f$  that gives identical results for similar
matrices.   Otherwise stated,  the results  given by  $f$ do  not depend  on the
reference  frame  into   which  the  matrices  are  expressed:   the  result  is
basis-invariant~\cite{Gantmacher1959TheorieMatricesV1,
  Koch1984Matrixinvariants}. For instance,  let $\Qb$ be a $m  \times m$ matrix,
and  $\Sb$  be any  invertible  matrix,  then  for  $f$ a  similarity  invariant
function, $f\left( \Sb  \, \Qb \, \Sb^{-1} \right) =  f\left( \Qb \right)$.  The
characteristic polynomial  $\det(\Qb-\lambda \Ib)$, where $\det$  stands for the
determinant, is a  similarity invariant, as are the $m-1$  coefficients $I_i$ of
the monomials $\lambda^{m-i-1}$.  These  latter are called principal invariants;
in the  case of a  $3 \times  3$ matrix, the  principal invariants are  given by
$I_0(\Qb)  = \Tr\Qb$,  $I_1(\Qb)  = \frac12  \left( \Tr  \Qb^2  - \left(  \Tr\Qb
\right)^2 \right)$ and $I_2(\Qb) = \det  \Qb$ with $\Tr$ the trace operator (see
Faddeev–LeVerrier
algorithm~\cite[IV-\S~5]{Gantmacher1959TheorieMatricesV1},~\cite{Hou1998ProofLeverrierFaddeev}
or~\cite[\S~3.8]{Spencer2004ContinuumMechanics}).

Applying these  definitions to the  magnetic gradient tensor $\Gb$  (remind that
$\Tr(\Gb)=0$) leads to
\begin{equation*}
\left\{
\begin{array}{ll}
I_0(\Qb) = 0 \\[2mm]
I_1(\Qb) = \frac12 \Tr \left(\Gb^2\right)\\[2mm]
I_2(\Qb) = \det(\Gb)
\end{array}
\right.
\end{equation*}

The values taken by these invariants do not depend on the coordinate system, and
in particular,  they can be  evaluated in the $x'y'z'-$coordinate  system, i.e.,
from  expression~\eqref{Eq:Gradient}  with  $\ub'_\gamma$ and  $\mb'$  replacing
$\ub_r$ and  $\mb$, respectively. The  results of this calculation  were already
obtained   in~\cite{Yan2024effective}  and   can  be   recovered  using   simple
algebra\footnote{For the  second case,  one may  judiciously apply  the rank-one
updates  determinant lemma  $\det\left( \Qb  + \alphab  \, \betab^\top\right)  =
\det(\Qb)   \left(  1   +  \betab^\top   \Qb   \,  \alphab   \right)$  and   the
Sherman–Morrison formula  $\left( \Qb  + \alphab  \, \betab^\top  \right)^{-1} =
\Qb^{-1}  -  \frac{\Qb^{-1} \alphab  \,  \betab^\top  \Qb^{-1}}{1 +  \betab^\top
  \Qb^{-1} \alphab} $~\cite[\S~6.2 \& \S~3.8 \&]{Meyer2200MatrixAnalysis}.}
\begin{subequations}
\begin{align}
\label{Eq:FirstInvariant}
I_1(\Gb) & = \frac{9 \, \left( \|  \mb' \|^2 + 2 \big( {\mb'}^\top \ub_r' \big)^2
  \right)}{r^8}\\[2.5mm]
\label{Eq:SecondInvariant}
I_2(\Gb) & = \frac{27 \,  \big( {\mb'}^\top \ub_r' \big) \left(
\| \mb' \|^2 + \big( {\mb'}^\top \ub_r' \big)^2 \right)}{r^{12}}
\end{align}
\end{subequations}

Consider now that  $\Pib(\Hb) = \frac12 \, \Tr\Gb^2$, i.e.,  the first invariant
is      recorded       along      the      sensor's       trajectory.      Using
Eqs.~\eqref{Eq:uPrime}-\eqref{Eq:rTrajectory}               and              the
expression~\eqref{Eq:FirstInvariant}, the noise free signal reads:
\begin{equation*}
\sb(u) =  \Ab \, \frac1{(1+u^2)^5} \: \begin{bmatrix}
1\\[1mm]
u\\[1mm]
u^2
\end{bmatrix}
\end{equation*}
with
\begin{equation}
\label{Eq:CoeffFirstGradInv}
\Ab = \frac{9}{D^8} \begin{bmatrix}
{m_x'}^2 + {m_y'}^2 + 3 \, {m_z'}^2  \\[2mm]
4 \, m_x' \, m_z'\\[2mm]
3 \, {m_x'}^2 + {m_y'}^2 + {m_z'}^2
\end{bmatrix}^\top
\end{equation}
As in  the previous situations, the  measured signals are decomposed  on a known
basis $\F$, now given by
\begin{equation*}
\label{Eq:BasisFirstGradInv}
\F = \left\{ u \mapsto \dfrac{u^i}{(1+u^2)^5} \right\}_{i=0}^2
\end{equation*}
Let us emphasize that the magnitude of the signal scales as $\left( \frac{\| \mb
  \|}{D^4} \right)^2$, making the use of  this signal much more sensitive to the
approaches based on $\Hb$.

For sake of  completeness, one may consider the case  where the second invariant
is measured and recorded along the sensor's trajectory. Following the same lines
as in  the previous  section, it  can be  shown that  the recorded  signal still
decomposes on  a basis,  and that  it scales  like $\left(\frac{\|  \mb \|}{D^4}
\right)^3$, exhibiting $\frac1{D^{12}}$ behaviour which makes it barely useable.
Details are deferred to the appendix~\ref{App:SecondInvariant}.

% ------------------------ Electromagnetic landscape

\subsection{Noise modeling}
\label{SubSec:NoiseModeling}

In fact, numerous sources of noise can distort the magnetic measurement. We will
assume that these can be reasonably  modeled as additive colored Gaussian random
disturbances that are statistically independent  of the anomaly signal. Anything
not attributable to the anomaly will be considered as contribution to the noise;
this includes the  Earth’s magnetic field variations,  sea swell, seabed-related
effects, the aircraft carrying the magnetometer,  and the intrinsic noise of the
magnetometer  itself.  Some  of  these  sources of  error  can be  significantly
reduced    using    appropriate    algorithms    (examples    can    be    found
in~\cite{Lelial1961Identification, Hezel2020ImprovingCalibration} or more recent
methods~\cite{Hezel2020ImprovingCalibration,  Nerrise2023physics}),  or, in  the
case  of airborne  measurements, unwanted  aircraft motions  can be  effectively
compensated  for  using  invariant  functions,  as  discussed  in  the  previous
section. Nevertheless, the diversity and  number of potential disturbances argue
in favor of the  centered colored Gaussian noise model by  virtue of the central
limit theorem  (it is assumed  that the mean cancels  out since it  is generally
known—for example, the  Earth’s field—or can be subtracted, for  example using a
reference sensor).

% ---------------------------- Detection Problem ----------------------------- %

\section{The detection problem}
\label{Sec:DetectionProblem}

% ------------------------ GLRT

\subsection{Notations and formulation of the test}
\label{SubSec:GLRT}

As outlined in section~\ref{SubSec:SignalModeling}, the  signal $\sb$ lives in a
$N$-dimensional space spanned by a basis  $\F$ of continuous functions ($N=3$ or
$N=4$  in  the examples  derived  in  the  preceding  section). In  the  sampled
digitized domain, this is  formulated as $\sb = \Ab \Fb$, where  each row of the
$N \times  K$ matrix $\Fb$ contains  $K$ samples spread over  the record length;
$\Ab \in  \Rset^{d \times N}$, contains  the expansion coefficients on  $\F$ for
each of  the $d$  measurement components (e.g.   $d=1$ for  scalar measurements,
$d=3$ for triaxial measurements). It is assumed  that $K > N$ is large enough to
warrant that  the elements of  the basis  $\F$ remain linearly  independent.  We
further  assume  that   the  noise  is  additive,   Gaussian  distributed,  with
independant spatial  and temporal correlations  represented by the $d  \times d$
positive definite matrix $\Sigmab$ and the $K \times K$ positive definite matrix
$\Psib$ respectively.  Thus, in the presence of a magnetic anomaly, the recorded
signal $\xb \in \Rset^{d \times N}$ reads
\begin{equation*}
\xb = \sb + \nb
\end{equation*}
where the  noise satisfies $  \nb \sim \N_{d,K}  \big( \zerob ,  \Sigmab \otimes
\Psib \big)$,  i.e., the noise  probability density  function (pdf) is  given by
$p_{\nb}(\xb)  \propto \exp  \left( -\frac12  \Tr\left( \Sigmab^{-1}  \, \xb  \,
\Psib^{-1}  \, \xb^\top  \right) \right)$  where the  normalization constant  is
skipped~\cite{Gupta2018matrix}.

The  MAD problem  can thus  be reformulated  as the  following classical  binary
hypothesis test:
\begin{equation*}
\left\{
\begin{array}{ll}
\H_0 \, : \, \xb \, = \, \nb       & \text{(absence of the source)}\\[1.5mm]
\H_1 \, : \, \xb \, = \, \sb + \nb & \text{(presence of the source)}\\
\end{array}
\right.
\raisetag{2\normalbaselineskip}
\end{equation*}

If  the  signal  $\sb$  is  known, the  classical  decision  strategy  for  this
test~\cite{Kay1998detection,  vanTrees2013DetectionEstimation} reads\footnote{$y
\overset{\H_1}{\underset{\H_0}{\gtrless}} \eta$ means that if $y > \eta$, $\H_1$
is decided, and conversely for $\H_0$.}
\begin{equation}
\label{Eq:LLRT-0}
\Lambda   =    \Big(   \L(\xb   |    \sb,   \H_1)-   \L(\xb   |    \H_0)   \Big)
\overset{\H_1}{\underset{\H_0}{\gtrless}} \eta
\end{equation}
where $\L =  \log p_{\xb}$ is the log-likelihood (respectively  under $\H_1$ and
given  $\sb$, and  under  $\H_0$). Note  that $\Lambda$,  referred  to as  ``the
receiver'',  does not  depend upon  the adopted  strategy or  the definition  of
optimality criterion for  the test; this dependance is translated  solely in the
values     chosen    for     the    threshold     $\eta$~\cite{Kay1998detection,
  vanTrees2013DetectionEstimation}.

Consider now  the ``square root factorization''  (e.g., Cholesky decompositions,
positive      definite     square-root,      etc.~\cite{Meyer2200MatrixAnalysis,
  Boyd2004ConvexOptimization,   Lasserre2015PolynomialOptimization})    of   the
positive definite covariance matrices $\Psib$ and $\Sigmab$:
\begin{equation}
\label{Eq:Cholesky}
\Psi^{-1} = \Lb_{\Psib} \, \Lb_{\Psib}^\top
,\quad   
\Sigmab^{-1} = \Lb_{\Sigmab} \, \Lb_{\Sigmab}^\top
\end{equation}
These factorizations  make it possible to  whiten the noise contribution  in the
signal       using      of       the      following       linear      transforms
(filtering)~\cite{Kay1998detection, vanTrees2013DetectionEstimation}
\begin{equation}
\label{Eq:FilteredSignal}
\txb = \Lb_{\Sigmab}^\top \, \xb \, \Lb_{\Psib}
,\qquad
\tsb = \Lb_{\Sigmab}^\top \, \sb \, \Lb_{\Psib}
,\qquad
\tnb = \Lb_{\Sigmab}^\top \, \nb \, \Lb_{\Psib}
\end{equation}
Finally, by  substituting the Gaussian  pdf into Eq~\eqref{Eq:LLRT-0},  the test
reduces here to
\begin{equation}
\label{Eq:LLRT}
\Lambda_c = 2 \Tr\left( \tsb \, \txb^\top \right)
-
\| \tsb \|^2_F \gtrless \eta
\end{equation}
with $\| \Qb \|_F = \sqrt{\Tr( \Qb \, \Qb^\top)}$ the Frobenius norm.

In the context  where $\sb$ depends on some unknown  parameters $\vartheta$, the
generalized  log-likelihood ratio  test  (GLRT) is  constructed,  for which  the
unknown  parameters are  replaced by  their maximum  likelihood estimator  (MLE)
$\hvarthetab$ in the expression of the receiver. Thus:
\begin{equation}
\label{Eq:GLLRT}
\Lambda = 2 \Tr\left( \tsb\big(\hvarthetab \big) \, \txb^\top \right)
-
\left\| \tsb\big( \hvarthetab \big) \right\|^2_F \gtrless \eta
\end{equation}
with
\begin{equation*}
\hvarthetab = \argmax_{\vartheta} \L\left( \xb | \sb,\H_1 \right)
\end{equation*}

% ------------------------ GLRT-Naive

\subsection{A straightforward approach}
\label{SubSec:GLRT-Naive}
\subsubsection{Derivation of the receiver}
\label{SubSubSec:NaiveReceiver}

Noting that  $\Fb$, $\Sigmab$  and $\Psib$  have been assumed  to be  known, the
parameters involved in the calculation of the receiver are the elements of $\Ab$
in the  expression $\sb = \Ab  \Fb$. After applying the  whitening transform, we
obtain $\tsb  = \Lb_{\Sigmab}^\top  \, \Ab  \, \Fb \,  \Lb_{\Psib}$ so  that the
filtered signal decomposes on the spatially filtered basis $\tFb$,
\begin{equation}
\label{Eq:CoefAFiltered}
\tsb = \tAb \, \tFb \qquad \mbox{with} \qquad \tFb = \Fb \, \Lb_{\Psib}, \quad
\tAb = \Lb_{\Sigmab}^\top \, \Ab
\end{equation}
The MLE of $\tAb$ is given by 
\begin{equation*}
\hAb_\F = \argmax_{\tAb \in  \Rset^{d \times N}} \, \L(\txb \, |  \, \tsb = \tAb
\, \tFb, \H_1)
\end{equation*}
or  equivalently (due  to the  Gaussian assumption,  which is  preserved through
linear transform)
\begin{equation*}
\hAb_\F = \argmin_{\tAb \in  \Rset^{d \times N}} \, \left\| \txb  - \tAb \, \tFb
\right\|_F^2
\end{equation*}
whose     solution     is     well     documented~\cite{Meyer2200MatrixAnalysis,
  Boyd2004ConvexOptimization, Magnus1999MatrixDifferential}.
    
By  introducing  the weighted  Frobenius  inner  product  and its  induced  norm
(hereafter called $\tFb$-inner product and $\tFb$-norm respectively):
\begin{equation}
\label{Eq:WeightedFrobenius}
\left\langle \Qb  \, , \,  \Rb \right\rangle_{\tFb} =  \Tr\left( \Qb \,  \tFb \,
\tFb^\top \, \Rb^\top \right), \quad \big\| \Qb \big\|_{\tFb}^2 = \left\| \Qb \,
\tFb \right\|_F^2
\end{equation}
the solution may be derived after some simple algebra, as\footnote{It comes from
$\left\| \txb - \tAb \, \tFb  \right\|_F^2 - \left\| \txb \right\|_F^2 + \left\|
\txb  \, \tFb^\top  \right\|_F^2  = \Tr\left(  \txb  \, \tFb^\top  -  2 \,  \txb
\tFb^\top \, \tAb^\top + \tAb \, \tFb \, \tFb^\top \, \tAb^\top \right)$.}
\begin{equation}
\hAb_\F =  \argmin_{\tAb \in \Rset^{d  \times N}} \,  \left\| \txb \,  \tFb^\# -
\tAb  \right\|_{\tFb}^2 ,  \:\: \tFb^\#  =  \tFb^\top \left(  \tFb \,  \tFb^\top
\right)^{-1}
\end{equation}
where     $\tFb^\#$      is     the     Moore-Penrose      pseudo-inverse     of
$\tFb$~\cite{Meyer2200MatrixAnalysis,     Magnus1999MatrixDifferential}.     The
immediate solution is thus trivially
\begin{equation}
\label{Eq:AHatTilde}
\hAb_\F = \txb \, \tFb^\#
\end{equation}
and using~\eqref{Eq:LLRT}, the test reads
\begin{equation}
\label{Eq:GLLRT-Naive}
\Lambda_\F = \left\| \txb \, \tFb^\# \right\|_{\tFb}^2 \gtrless \eta
\end{equation}

One of  the advantages of  such a detector  is its linearity,  which facilitates
calculations   and  allows   for  analytical   derivations.   Furthermore,   the
performance of  this simple proposed  detector (named ``naive'' receiver  in the
sequel)~\eqref{Eq:GLLRT-Naive}  can  be easily  computed,  as  explained in  the
following subsection.
 
\

% ------------------------ Performance

\subsubsection{Receiver performance}
\label{SubSubSec:Performance}

The detector's  performance is evaluated  by calculating the  Receiver Operating
Characteristic (ROC), which expresses the  probability of detection $\Pd(\eta) =
\Pr\left[ \Lambda  > \eta |  \H_1 \right]$ as a  function of the  probability of
false  alarm  $\Pfa(\eta)   =  \Pr\left[  \Lambda  >  \eta   |  \H_0\right]$.  A
comprehensive  study  of  the  --widely  used--  ROC  curves  can  be  found  in
e.g.,~\cite{Kay1998detection,  vanTrees2013DetectionEstimation}.  Let us  simply
keep in mind that the ROC must be concave everywhere (otherwise a better test is
easily constructed), that the ROC is  parametrized by the threshold $\eta$ (with
fixed points  $\left( \Pfa  , \Pd  \right) =  (0,0)$ for  $\eta =  -\infty$) and
$\left( \Pfa , \Pd \right) = (1,1)$  for $\eta = +\infty$), and finally that the
lower the $\Pfa$ and the higher $\Pd$, the better the receiver.

Once the pdf of  the receiver has been identified, the ROC  curves can be easily
obtained:
\begin{equation}
\label{Eq:Perf}
\left\{\begin{array}{lll}  
\Pfa(\eta) = \ccdf_{\Lambda | \H_0}(\eta) \\[4mm]
\Pd(\eta)= \ccdf_{\Lambda | \H_1}(\eta)
\end{array}\right.
\end{equation}
where $\ccdf = 1-\cdf$ denotes the complementary cumulative density function.

When $\sb$ is perfectly known, the receiver calculated in Eq.~\eqref{Eq:LLRT} is
called  ``clear-seeing'' and  takes the  expression  here $\Lambda_c  = \|  \tsb
\|_F^2 + 2 \,  \Tr\left( \tsb \, \tnb^\top \right)$.  $\tnb$  is a random matrix
with  standard independent  Gaussian components,  as is  $\Lambda_c$ (by  linear
combination  of these  components, plus  a constant),  with mean  $k \,  \| \tsb
\|_F^2$ under $\H_k$, and variance $4 \, \| \tsb \|_F^2$
\begin{equation}
\label{Eq:StatClairvoyant}
\Lambda_c | \H_k \sim \N\left(  k \, \| \tsb \|_F^2 \, , \,  4 \, \| \tsb \|_F^2
\right)
\end{equation}

In   order   to   derive   the   performance   of   the   ``naive''   receiver's
in~\eqref{Eq:GLLRT-Naive}, let  us set  $\left( \tFb \,  \tFb^\top\right)^{-1} =
\Lb_{\tFb} \Lb_{\tFb}^\top$ (e.g., by Cholesky decomposition).
%\footnote{In         general,        the         basis        $\tFb$         is
%orthonormalized~\cite{Chenevas2024analytical,             Sheinker2009magnetic,
%Ginzburg2002processing},  for   instance  by  a  Gram-Schmidt   or  Householder
%procedure~\cite{Meyer2200MatrixAnalysis},  which  is   precisely  the  case  of
%$\Lb_{\tFb}^\top \, \tFb$.}.
Using   definition~\eqref{Eq:WeightedFrobenius},   a    few   simple   algebraic
calculations  allow  us to  express  $\Lambda_\F  =  \left\| \txb  \,  \tFb^\top
\Lb_{\tFb} \right\|_F^2$, a quantity for which the pdf must be calculated. As by
construction,  the matrix  $\tFb^\top \Lb_{\tFb}  $ satisfies  $\left( \tFb^\top
\Lb_{\tFb}  \right)^T \tFb^\top  \Lb_{\tFb}  = \Ib$,  it  has orthonormal  rows;
therefore,  the noise  contribution  in  $\txb$ (remind  that  $\tnb$ is  white,
according to section~\ref{SubSec:GLRT}) is  Gaussian (only linear transforms are
involved) and  whiteness is  preserved.  As a  consequence, for  each hypothesis
$\H_k, \: k = 0, \, 1$,
\begin{equation*}
\left.  \txb \,  \tFb^\top \, \Lb_{\tFb} \right| \H_k \,  \sim \N_{d,N} \left( k
\, \tsb , \Ib_d \otimes \Ib_N \right)
\end{equation*}
This allows to conclude that the receiver follows a chi-square distribution with
degree of freedom $\nu = d \, N$ and non-centrality parameter $\lambda = k \, \|
\tsb \|_F^2$~\cite{Johnson1995:v1, Patnaik1949NonCentral}:
\begin{equation}
\label{Eq:StatNaive}
\Lambda_\F | \H_k \sim \chi^2_{d N}\left( k \, \| \tsb \|_F^2 \right)
\end{equation}

Remark: Since  $\| \tsb  \|_F^2 =  \| \Lb_{\Sigmab}^\top  \, \sb  \, \Lb_{\Psib}
\|_F^2$, the non-centrality parameter has the meaning of a signal-to-noise ratio
(SNR), up to a factor of $1/K$: it is clear that the performance of the receiver
is an  increasing function of  the SNR  as it is  the case for  the clear-seeing
receiver.

%Examples are presented and discussed in section \ref{Sec:Simulation}. 

% ----------------------- Physically constrained GLRT ------------------------- %

\section{Physics\--aware generalized log-likelihood ratio test}
\label{Sec:PhysicallyConstrainedGLRT}

In  the  previous  section,   the  receiver~\eqref{Eq:GLLRT-Naive}  was  derived
regardless of the fact  that the space spanned by $\tFb$  may contain subsets of
vector values that have no physical reality.  In reality, the coefficients $\Ab$
describing the signal as an expansion  on the vector basis $\tFb$ are structured
(or constrained) by the physical form of the magnetic dipole radiation field, as
expressed             by             equations~\eqref{Eq:CoeffProjectedAnomaly},
\eqref{Eq:CoeffSquareModulus},                      \eqref{Eq:CoeffFirstGradInv}
or~\eqref{Eq:CoeffSecondGradInv}  respectively,  depending  on  the  measurement
strategy adopted. Similarly, these constraints are propagated to $\tAb$ when the
transformation~\eqref{Eq:CoefAFiltered} is applied. In the sequel, the following
notations are  adopted: $\varthetab$ is  the vector of parameters,  of dimension
$\nu$, varying  in the space $\Theta$  (for example, $\varthetab =  \mb$, $\nu =
3$, $\Theta = \Rset^3)$.

Notice  now  that $\tAb  \in  \Rset^{d  \times  N}$  is parameterized  by  $\nu$
parameters:  it implies  so  that when  $\nu  < d  \, N$,  $\tAb$  belongs to  a
semi-algebraic sub-space $\tV \subsetneq \Rset^{d \times N}$ of lower dimension.
Although  such simple  dimensional argument  do not  hold for  scalar (including
square-modulus) measures, we can still have $\tV \subsetneq \Rset^{1 \times 3}$,
because the physical constraints outlined above must always be satisfied.

Therefore, the  proposed approach consists in calculating the MLE of  $\tAb$ by
reducing the search space to the constrained space $\tV$, leading to reformulate
Eq.~\eqref{Eq:GLLRT} as
\begin{equation}
\label{Eq:GLLRT-V}
\Lambda_\V =  2 \, \left\langle \hAb_\V  \, , \, \hAb_\F  \right\rangle_{\tFb} -
\left\| \hAb_\V \right\|^2_{\tFb} \gtrless \eta
\end{equation}
where $\hAb_\F =  \xb \, \tFb \,  \left( \tFb \, \tFb^\top  \right)^{-1}$ is the
maximum likelihood estimator in the unconstrained space~\eqref{Eq:AHatTilde} and
where
\begin{subequations}
\label{Eq:Ah}
\begin{align}
\label{Eq:AhV}
\hAb_\V  & =  \argmin_{\tAb \in  \tV} \left\|  \hAb_\F -  \tAb \right\|_{\tFb}^2
\\[2mm]
\label{Eq:AhTheta}
& =\tAb(\hvarthetab)  \quad \mbox{with} \quad \hvarthetab  = \argmin_{\varthetab
  \in \Theta} \left\| \hAb_\F - \tAb(\varthetab) \right\|_{\tFb}^2
\end{align}
\end{subequations}
%
%Thus,  an alternate  equivalent approach  emerges, based  on the  estimation of
%$\hvarthetab$.
Indeed,   simple   calculations   allow    to   show   that   $\hvarthetab$   in
Eq.~\eqref{Eq:AhTheta} is obtained for $\hAb = \tAb(\hvarthetab)$, solution of

\begin{equation}
\label{Eq:ThetaOpt}
\begin{array}{cc}
\forall i = 1,  \ldots , \nu , & \displaystyle \left\langle \hAb_\F  - \tAb \: ,
\: \pdv{\tAb^\top}{\vartheta_i} \right\rangle_{\tFb} = 0
\end{array}
\end{equation}
where  $\vartheta_i$  is the  $i$-th  component  of $\varthetab$.   Finally,  if
$\hAb_\F \in \tV = \tAb(\Theta)$, the trivial solution of~\eqref{Eq:AhTheta} has
a  physical relevance,  and  $\hAb_\F$ is  accepted as  an  estimate of  $\tAb$.
Otherwise,  according to~\eqref{Eq:ThetaOpt}  the  physically admissible  better
estimate $\hAb$ is  the orthogonal projection of $\hAb_\F$ on  the tangent space
of  the  boundary $\partial  \tV$  of  $\tV$~: the  error  $\hAb_\F  - \hAb$  is
orthogonal\footnote{Orthogonality  is  to  be  understood with  respect  to  the
$\tFb$-inner  product; thus  it is  an oblique  projection wrt  the usual  inner
product, as  a consequence of the  non orthonormality of basis  $\tFb$.}  to the
tangent  space of  $\partial\tV$  at  $\hAb$ (assuming  that  the tangent  space
exists).

Note that the identical solution appears when applying the orthogonal projection
theorem to Eq.~\eqref{Eq:AhV}:  First, an estimate is  constructed by projecting
the observation on the basis, then this latter estimate is projected on $\tV$ if
necessary.  Either  of these  two approaches  can be  applied, depending  on the
setting and their ability to lead to simple implementations.

{\em  Remark}: Additional  constraints may  be  introduced, for  example on  the
magnetic  dipole  itself  as  in~\cite{Yang2018magnetic}  in  a  scalar  measure
framework, which again  leads to constraining the coefficient space.   As in the
context of  this paper, no  assumptions are made  about the distribution  of the
dipole magnetic moment in $\Rset^3$, this will not be studied in further detail.

% ------------------------ Formalization

\subsection{$\tV$ is a semi-algebraic space}
\label{SubSec:Formal}

Extanding Eq.~\eqref{Eq:CoefAFiltered}  to all physically  acceptable solutions,
we get
\begin{equation}
\tV =  \Lb_{\Sigmab}^\top \, \V \quad \mbox{with} \quad \V = \Ab(\Theta)
\end{equation}
By consequence,  identifying $\V$ is  sufficient to recover  $\tV$.  Determining
the       nature       of        $\V$       relies       on       semi-algebraic
geometry~\cite{Coste2000introduction,  Lasserre2015PolynomialOptimization}.   In
the  context of  relation~\eqref{Eq:CoeffProjectedAnomaly}, consider  the angles
$\alpha$ and $\beta$ introduced in section~\ref{SubSec:Geometry}; they are fully
characterized by the pairs  $(c_\alpha,s_\alpha) = (\cos\alpha, \sin\alpha)$ and
$(c_\beta,s_\beta)  =  (\cos\beta,  \sin\beta)$  respectively  and  satisfy  the
trigonometric constraints $c_\alpha^2 + s_\alpha^2  = 1$, $c_\beta^2 + s_\beta^2
=  1$.    The  parameter   vector  to  consider   is  therefore   $\varthetab  =
(\mb,c_\alpha,s_\alpha,c_\beta,s_\beta)  \in \Rset^7$,  ($\nu  = 7$).   Overall,
$(\Ab,\varthetab)$ defines in turn a $(d N + \nu)-$dimensional space constrained
by polynomial  relations denoted  generically $\P(\varthetab)$,  which precisely
corresponds  to  the  definition  of  a  semi-algebraic  space  of  $\Rset^{d  N
  +\nu}$~\cite{Coste2000introduction,  Lasserre2015PolynomialOptimization}. $\V$
is the projection of this space onto the $d N$ first variables and, according to
the Tarski-Seidenberg theorem~\cite{Coste2000introduction},  is a semi-algebraic
space.

Thus $\V$ is rewritten as
\begin{equation*}
\V =  \left\{ \Ab \in \Rset^{d  \times N} \:  \big| \: \exists \,  \varthetab \:
\mbox{ s. t. } \: \P(\varthetab) \right\}
\end{equation*}
where  $\exists$ is  called  {\em  existential quantifier};  this  space can  be
identified        using        a       {\em        quantifier        elimination
  algorithm}~\cite{Coste2000introduction,    Lasserre2015PolynomialOptimization,
  Sturm2017SurveyQuantifierElimination},         e.g.          using         the
package~\cite{Dolzmann1997redlog}   when  possible   (since  the   computational
complexity   is    at   least    doubly   exponential    in   the    number   of
quantifiers~\cite{Basu1994ComplexityQuantifier,
  Davenport1988QuantifierExponential}).
% and  in the  number  of variables  with  base  the product  of  the number  of
% polynomials and their maximum degree

In  order to  deepen the  analysis,  $\V$ must  be  identified for  each of  the
scenarios presented in the  preceding sections.  Whenever $\V \subsetneq\Rset^{d
  \times  N}$, there  exist  signals in  $\vect  \tFb$ that  are  not in  $\tV$.
Although  such  occurences  have  no  physical  reality,  the  naive  method  of
subsection~\ref{SubSec:GLRT-Naive} did not exclude them.

In   the  following   subsections,   the  scenario   and   signals  studied   in
subsection~\ref{SubSec:SignalModeling} are reexamined; the associated estimators
$\hAb$  Eqs.~\eqref{Eq:Ah} are  calculated, as  well as  the corresponding  {\em
  constrained} GLRT~\eqref{Eq:GLLRT-V}.   It should  be noted  that $\tAb$  is a
homogeneous  polynomial in  the  variables $(m_x',m_y',m_z')$,  whose degree  is
equal to one for the $d$-axis magnetometer and for the scalar sensor. The degree
of the  polynomial is equal  to two for the  square modulus measurement  and the
first invariant measurement, while it is equal to three for the second invariant
measurement.   In what  follows, the  focus is  on the  case of  the linear  and
quadratic polynomial, for which practical solutions are exhibited.

% ------------------------ d-Axis case

\subsection{Homogeneous linear case: the $d-$axis magnetometer}
\label{SubSec:dAxisMagnometer}

\subsubsection{The case $d=1$}

In the single axis measurement case,  $\Ab \in \Rset^3$ is parameterized by $\nu
= 4$ or  $5$ parameters. The identification of $\Ab$  is thus not overdetermined
and  does not  allow us  to  conclude about  a possible  dependence between  the
coefficients.    Denoting   $\Pib'  =   \begin{bmatrix}   \pi'_x   &  \pi'_y   &
  \pi'_z                                                    \end{bmatrix}^\top$,
Eqs.~\eqref{Eq:MMatrixdAxis}-\eqref{Eq:CoeffProjectedAnomalyPip}      can     be
rewritten as
\begin{equation}
\Ab = {\mb'}^\top \Pb' \qquad \mbox{with} \qquad \Pb' = \begin{bmatrix}
- \pi'_x    & 3 \, \pi'_z & 2 \, \pi'_x\\[2mm]
- \pi'_y    &      0      &    - \pi'_y\\[2mm]
2 \, \pi'_z & 3 \, \pi'_x &    - \pi'_z\\[2mm]
\end{bmatrix}
\end{equation}
The determinant of $\Pb'$ can easily be  calculated and is given by $\det \Pb' =
-9 \, \pi'_y \left( {\pi'_x}^2 + {\pi'_z}^2 \right)$.  This leads to:
\begin{equation}
\Ab(\Theta) \neq \Rset^{1 \times 3} \: \Leftrightarrow \: \pi'_x = \pi'_z = 0 \,
\lor \pi'_y = 0
\end{equation}

Assuming $\pi'_x  = \pi'_z = 0$  corresponds to assume that  the magnetic sensor
(or the Earth's magnetic field) is perfectly aligned with the $y'-$axis (defined
by the vector source-sensor).  The probability of this occurring is zero.

However $\pi'_y = 0$ may happen, for  instance in the situation where the sensor
axis  and the  trajectory are  aligned,  or in  the  case of  the scalar  sensor
described  sub-subsection~\ref{SubSubSec:ScalarMeasurement},  when the  sensor's
trajectory is  parallel to the Earth's  field lines.  Apart from  these specific
exceptions,  that are  extremely difficult  to  satisfy in  practice, the  naive
approach of subsection~\ref{SubSec:GLRT-Naive} cannot be improved.

\

\subsubsection{The case $d > 1$}
\label{SubSec:ThreeAxisPerformance}

This case  is characterized by  $\nu < 3  \, d $  (whether $\alpha$ is  known or
not),  which  imposes $\V  \neq  \Rset^{d  \times  3}$, being  a  semi-algebraic
sub-space of $\Rset^{d \times 3}$.

The  semi-algebraic  space $\V$  is  difficult  to  identify directly;  in  this
section, a direct maximum likelihood  estimation of the physical coefficients is
presented.

Such an approach  was initiated in~\cite{Chenevasdetection}, for the  case of an
airborne vector magnetometer  ($\alpha=0$) in additive Gaussian  white noise. It
is extended  in this  section to a  more general framework  for fixed  or moving
sensor in additive colored noise.

From
Eqs.~\eqref{Eq:AhTheta}-\eqref{Eq:CoefAFiltered}-\eqref{Eq:CoeffProjectedAnomalyPip}
the parameter vector $\varthetab =  (\mb',\alpha,\beta)$ must be estimated, as a
solution of the minimization problem
\begin{equation*}
\min_{\alpha,\beta} \min_{\mb'}\left\| \txb - \Lb_{\Sigmab}^\top \, {\Pib'}^\top
\, \Mb' \, \tFb \right\|_F^2
\end{equation*}
This  minimization can  be  performed as  a {\em  two-stage}  or {\em  bi-level}
minimization                            problem~\cite{Colson2007Overviewbilevel,
  Dempe2002Bilevelprogramming}.

Firstly,  the  {\em  inner}  minimization,  wrt $\mb'$,  is  reformulated  as  a
classical linear  regression problem; actually,  for any matrices  $\Pb$, $\Qb$,
$\Rb$ (with compatible  dimensions so that their product  exists), the following
equalities hold~\cite[Chap.~2, Sec.~4]{Magnus1999MatrixDifferential}:,
\begin{eqnarray*}
\Tr \left( \Pb^\top \Qb  \right) & = & \vec(\Pb)^\top \vec(\Qb),\\[2mm]
\vec\left( \Pb \,  \Qb \, \Rb\right) &  = & \left( \Rb^\top  \otimes \Pb \right)
\vec \Qb
\end{eqnarray*}
where    $\vec$   denotes    the   vectorization    operator\footnote{Let   $\Pb
= \begin{bmatrix}  \pb_1 &  \ldots &  \pb_n \end{bmatrix}$  a $m  \times n$-size
matrix;    $\vec(\Pb)    =    \begin{bmatrix}     \pb_1^\top    &    \ldots    &
  \pb_n^\top \end{bmatrix}^\top$ is the $n m \times 1$ column vector obtained by
concatenating the columns  of $\Pb$.}.  Therefore, the  above inner minimization
problem can be reformulated as
\begin{equation*}
\hmb  = \argmin_{\mb'  \in \Rset^3}  \left\|  \vec\left( \txb  \right) -  \left(
\tFb^\top  \otimes \left(  \Lb_{\Sigmab}^\top  \,  {\Pib'}^\top \right)  \right)
\vec\left( \Mb' \right) \right\|^2
\end{equation*}
From~\eqref{Eq:MKroneckerdAxis} and since  the $3  \times 3$  blocs of
$\Cb$ are  symmetrical
% By noticing\footnote{This is  due to the fact  that the $3 \times  3$ blocs of
% $\Cb$ are symmetrical.} that (see \eqref{Eq:MKroneckerdAxis})
$\vec\left( \Mb'  \right) =  \vec\left( \Cb  \, \big(  \Ib_3 \otimes  \mb' \big)
\right) = \Cb^\top \, \mb'$, we get the least square problem
\begin{equation}
\hmb =  \argmin_{\mb' \in  \Rset^3} \left\| \vec\left(  \txb \right)  - \Rb^\top
\mb' \right\|^2
\label{mLSE}
\end{equation}
where
\begin{equation}
\Rb = \Cb \left( \tFb \otimes \left( \Pib' \, \Lb_{\Sigmab} \right) \right)
\end{equation}
The      solution      of      \eqref{mLSE}     is      well      known      and
reads \cite{Meyer2200MatrixAnalysis,                  Boyd2004ConvexOptimization,
  Magnus1999MatrixDifferential}
\begin{equation}
\hmb = {\Rb^\#}^\top \, \vec\left( \txb \right), \qquad \Rb^\# = \Rb^\top \left(
\Rb \, \Rb^\top \right)^{-1}
\end{equation}
where     $\Rb^\#$    stands     for     the     Moore-Penrose    inverse     of
$\Rb$~\cite{Meyer2200MatrixAnalysis, Magnus1999MatrixDifferential}.

Secondly, the expression obtained for $\hmb$ is  used in order to solve the {\em
  outer} optimization problem, that reads:
\begin{equation}
\min_{\alpha,\beta} f, \qquad  f = \left\| \left( \Ib -  \Rb^\# \Rb \right)
\vec\left( \txb \right) \right\|^2
\end{equation}
The  symmetrical  matrix   $\Rb^\#  \Rb  =  \Rb^\top  \left(   \Rb  \,  \Rb^\top
\right)^{-1}  \Rb$ is  nothing but  the  orthogonal projection  matrix onto  the
subspace generated by the rows of $\Rb$; thus an alternate formula for $f$ is
\begin{equation}
f  =  \vec\left(  \txb  \right)^\top  \left( \Ib_{d  K}  -  \Rb^\#  \Rb  \right)
\vec\left( \txb \right)
\end{equation}
Therefore,  the values  requested  for  $\alpha$, $\beta$  are  the values  that
minimize the norm of the projection error. Unfortunately, analytical derivations
to   solve  this   non  convex   optimization  problem   seem  to   be  out   of
reach. Nevertheless, since most  practical detection scenario requires real-time
processing, the choice of the optimization algorithm is a critical issue; global
optimization algorithms  are far too slow,  so we are turning  to gradient-based
methods. These  methods can take  advantage of the fact  that it is  possible to
obtain  an  analytical  expression  for  the gradient  of  the  function  to  be
minimized, in order to speed up  numerical optimization steps. After a few lines
of    calculation,    the    gradient   $\begin{bmatrix}    \pdv{f}{\alpha}    &
  \pdv{f}{\beta}\end{bmatrix}^\top$    is    shown   to    satisify\footnote{The
calculation  relies  on  the   equality  $\pdv{\Qb^{-1}}{\theta}  =  -  \Qb^{-1}
\pdv{\Qb}{\theta}                \Qb^{-1}$~\cite[Chap.~9,               Sec.~13,
  Eq.~(18)]{Magnus1999MatrixDifferential}, for any  invertible matrix $\Qb$, and
on the property $\vb^\top  \, \Qb \, \vb = \vb^\top \, \Qb^\top  \, \vb$ for any
vector $\vb$ and matrix $\Qb$.}:
\begin{equation}
\label{Eq:Gradient_f}
\pdv{f}{\theta}  =  2 \vec\left(\xb\right)^\top \, \Rb^\# \pdv{\Rb}{\theta}
\left( \Rb^\# \Rb^\top - \Ib_{d K} \right)^\top
\vec\left(\xb\right)
\end{equation}
where $\theta$ refers either to  $\alpha$ or to $\beta$.

In   practice,  Limited-memory   Broyden–Fletcher–Goldfarb–Shannon  quasi-Newton
algorithm    with     bound    constraints    (L-BFGS-B)~\cite{Zhu1997algorithm,
  Byrd1995limited} is  used in  the simulation  section~\ref{Sec:Simulation}. It
implements an approximation of the Hessian matrix based on gradient evaluations,
which    is    very    effective    since   the    gradient    is    given    by
Eq.~\eqref{Eq:Gradient_f}.   Note  that  as  with  all  quasi-Newton  algorithm,
convergence to the  global minimum is not guaranteed, and  the solution obtained
depends on the initialization.  In addition to these theoretical considerations,
L-BFGS-B is widely implemented in standard scientific computing libraries across
multiple  programming languages,  facilitating  its  utilization.  In  practice,
L-BFGS-B  offers a  favorable  compromise between  computational efficiency  and
convergence robustness.

% ------------------------ Square modulus performance

\subsection{Homogeneous quadratic situation: Square modulus of the anomaly and first invariant}
\label{SubSec:SquarePerformance}

Since  $d  =  1$  for  both square  modulus  or  first  invariant  measurements,
$\Lb_{\Sigmab}   \in  \Rset_+$   and  without   loss  of   generality,  we   set
$\Lb_{\Sigmab} = 1$ (the constant is absorbed in the temporal covariance). Thus,
$\tAb =  \Ab$ and $\tV  = \V$.  Furthermore, in  the specific framework  of this
study (see eq.  \eqref{Eq:CoeffSquareModulus} and \eqref{Eq:CoeffFirstGradInv}),
$\Ab$ generically reads:
\begin{equation}
\label{Eq:AbSquareInvGeneric}
\Ab \propto \begin{bmatrix}
{m_x'}^2 + {m_y'}^2 + a \, {m_z'}^2  \\[2mm]
2 \, b \, m_x' \, m_z'\\[2mm]
a \, {m_x'}^2 + {m_y'}^2 + {m_z'}^2
\end{bmatrix}^\top
\end{equation}
with  $a  >  1,  \:  b  >  0$.   As  in  the  case  of  the  single-axis  sensor
(subsection~\ref{SubSec:dAxisMagnometer}), the problem is not overdetermined and
no conclusion or constraints can be drawn a priori.

A direct optimization on $\mb$ could  be considered: As $\Ab \in \Rset^{1 \times
  3}$, the  set of equations~\eqref{Eq:ThetaOpt}  leads to estimating  $\mb'$ by
solving
\begin{equation}
\left( \hAb_\F - \Ab\left(\mb'\right) \right) \, \Fb  \, \Fb^\top \,
\begin{bmatrix*}[c]
m_x'  &  m_y'  &  a \, m_z'  \\[1mm]
 b \, m_z'  &  0  & b \, m_x'\\[1mm]
a \, m_x'  &  m_y'  &  m_z'
\end{bmatrix*} = \zerob^\top
\end{equation}
%
%(the argument of  the trace is a scalar).
The equality  above is equivalent  to three  equations satisfied by  three cubic
polynomials  in three  variables,  which prevents  this  approach from  yielding
simple analytic solutions.

\

We thus turn  to the semi-algebra geometrical approach by  identifying $\V$, and
show that $\V = \Ab(\Theta) \ne \Rset^{1  \times 3}$.  The form of $\Ab$ already
shows  simply that  $\V$  is restricted  to  row vectors  whose  first and  last
components are  positive. But  the restriction is  actually stronger;  using the
existential    quantifier     elimination    algorithm~\cite{Dolzmann1997redlog,
  Sturm2017SurveyQuantifierElimination},  or a  direct  approach,  the space  of
physically   feasible  coefficients   can  be   formally  identified   (detailed
explanation is provided in the appendix~\ref{App:HalfCone}), and reads:
\begin{equation}
\V = \left\{ \Ab \in \Rset^{1 \times 3} \: \big| \: \Ab \, \Qb_{a,b} \, \Ab^\top
\leqslant 0 \, \land \, \Ab \, \one_z \geqslant 0 \right\}
\label{Eq:Cone}
\end{equation}
where
\begin{equation*}
\Qb_{a,b} = \begin{bmatrix*}[c]
4 \, a \, b^2 & 0 & - 2 \, b^2 \left( a^2 + 1 \right)\\[2.5mm]
 0 & \left( a^2 - 1 \right)^2 & 0\\[2.5mm]
- 2 \, b^2 \left( a^2 + 1 \right) & 0 & 4 \, a \, b^2
\end{bmatrix*} 
\end{equation*}
and $\one_z = \begin{bmatrix} 0 & 0 & 1\end{bmatrix}^\top$.

Note that, $\Qb_{a,b}$ can be written\footnote{$\diag\vb$ is the diagonal matrix
whose diagonal terms are the components of $\vb$.} as
\begin{equation*}
\Qb_{a,b}  = \Rb_y\left( \frac{\pi}4  \right)  \,  \diag \begin{bmatrix*}[c]
  2 \,   b^2  \left( a + 1 \right)^2 \\[3mm]
       \left( a^2 - 1 \right)^2      \\[3mm]
- 2 \, b^2 \left( a - 1 \right)^2
\end{bmatrix*} \, \Rb_y\left( \frac{\pi}4 \right)^\top
\end{equation*}
where
$\Rb_y\left( \frac{\pi}4 \right) =  \frac{\sqrt2}2 \begin{bmatrix*}[c]
 1  &   0    & 1 \\[1mm]
 0  & \sqrt2 & 0 \\[1mm]
-1  &   0    & 1
\end{bmatrix*}$
is the rotation matrix of angle $\frac{\pi}4$ about the $y-$axis.

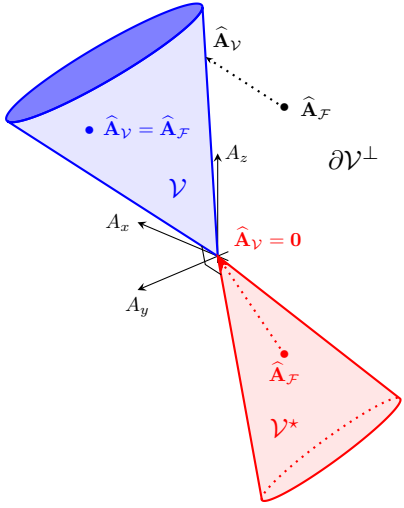
\begin{figure}[ht]
\begin{center}
\begin{tikzpicture}
    
%% some definitions
\pgfmathsetmacro{\El}{25} % elevation
\pgfmathsetmacro{\Az}{135}% azimuth

% Projection to plot a point (x,y,x) wrt angle of vision
\pgfmathdeclarefunction{projx}{2}{\pgfmathparse{#1*cos(\Az)-#2*sin(\Az)}}
\pgfmathdeclarefunction{projy}{3}{\pgfmathparse{#1*sin(\Az)*sin(\El)+#2*cos(\Az)*sin(\El)+#3*cos(\El)}}

\pgfmathsetmacro\a{3} % square modulus a = 4, first invariant a = 3
\pgfmathsetmacro\b{2} % square modulus b = 3, first invariant b = 2
\pgfmathsetmacro\z{3} % z for the plot of the elliptical basis
\pgfmathsetmacro\da{(\a-1)*\z/(\a+1)} % semi x-axis V
\pgfmathsetmacro\db{sqrt(2)*\b*\z/(\a+1)} % semi y-axis V
%
% schematic, Orth. basis and factor
\pgfmathsetmacro\das{.25*(\a+1)*\z/(\a-1)} % semi x-axis V^perp
\pgfmathsetmacro\dbs{.25*(\a+1)*\z/\b/sqrt(2)} % semi y-axis V^pert

% Axes plots
% ----------
%
\draw[>=stealth,  ->]
({projx(-.2,0)},{projy(-.2,0,0)})--({projx(1.5,0)},{projy(1.5,0,0)})
node[left,scale=.75]{$A_x$};
\draw[>=stealth, ->]
({projx(0,-.2)},{projy(0,-.2,0})--({projx(0,1.5)},{projy(0,1.5,0)})
node[below,scale=.75]{$A_y$};
\draw[>=stealth, ->]
(0,{projy(0,0,-.1)})--(0,{projy(0,0,1.5)})
node[right,scale=.75]{$A_z$};

% Cone V with its elliptical base
% -------------------------------
%
\filldraw[thick, color=blue, domain=55:225, variable=\t, samples=100, fill opacity=.1] (0,0)--(
     {projx(-\da*cos(55)+\z,-\db*sin(55))/sqrt(2)},
     {projy(-\da*cos(55)+\z,-\db*sin(55),\da*cos(55)+\z)/sqrt(2)})-- plot(
     {projx(-\da*cos(\t)+\z,-\db*sin(\t))/sqrt(2)},
     {projy(-\da*cos(\t)+\z,-\db*sin(\t),\da*cos(\t)+\z)/sqrt(2)})--cycle;
\filldraw[thick, color=blue, domain=0:360, variable=\t, samples=200, fill opacity=.5] plot(
     {projx(\da*cos(\t)+\z,\db*sin(\t))/sqrt(2)},
     {projy(\da*cos(\t)+\z,\db*sin(\t),-\da*cos(\t)+\z)/sqrt(2)});
\node[color=blue, scale=1] at ({projx(\z/4,0)},{projy(\z/4,0,\z/4}) {$\V$};

% Cone V^* with its elliptical base (case F ortho ; it's schematic)
% -----------------------------------------------------------------
%
\filldraw[thick, color=red, domain=30:210, variable=\t, samples=100, fill opacity=.1] (0,0)--(
     {projx(-\das*cos(30)-\z,-\dbs*sin(30))/sqrt(2)},
     {projy(-\das*cos(30)-\z,-\dbs*sin(30),\das*cos(30)-\z)/sqrt(2)}) -- plot(
     {projx(-\das*cos(\t)-\z,-\dbs*sin(\t))/sqrt(2)},
     {projy(-\das*cos(\t)-\z,-\dbs*sin(\t),\das*cos(\t)-\z)/sqrt(2)})--cycle;
\draw[thick, dotted, color=red, domain=-330:-140, variable=\t, samples=100] plot(
     {projx(\das*cos(\t)-\z,\dbs*sin(\t))/sqrt(2)},
     {projy(\das*cos(\t)-\z,\dbs*sin(\t),-\das*cos(\t)-\z)/sqrt(2)});

\node[color=red, scale=1] at ({projx(0,-.42*\z)},{projy(0,-.42*\z,-.975*\z}) {$\V^\star$};

% V^perp
% ------
%
\node[scale=1] at ({projx(0,-.85*\z)},{projy(0,-.85*\z,.2*\z}) {$\partial\V^\perp$};

% Right angle
% -----------
%
\coordinate (A) at (
  {projx(\da/sqrt(2)+\z,\db/sqrt(2))/sqrt(2)},
  {projy(\da/sqrt(2)+\z,\db/sqrt(2),-\da/sqrt(2)+\z)/sqrt(2)}
  );
\coordinate (B) at (0,0);
\coordinate (C) at (
  {projx(\das/sqrt(2)-\z,\dbs/sqrt(2))/sqrt(2)},
  {projy(\das/sqrt(2)-\z,\dbs/sqrt(2),-\das/sqrt(2)-\z)/sqrt(2)}
  );
\draw pic[draw, scale=.5]{right angle=A--B--C};

% A_F inside V
% -------------
%
\node[color=blue, scale=.75] at ({projx(1.5,0)},{projy(1.5,0,1.4)}) {$\bullet \:\: \hAb_\V = \hAb_\F$};

% A_F inside V^*
% ---------------
%
\draw[dotted, thick, -{Stealth[length=2mm]}, color=red] ({projx(0,-1.25)},{projy(0,-1.25,-1.85)})
     node[scale=.75] {$\bullet$} --
(0,0)     node[above right, scale=.75] {$\:\:\hAb_\V = \zerob$};
\draw[color=red] ({projx(0,-1.25)},{projy(0,-1.25,-1.85)}) node[below, scale=.75] {$\hAb_\F$};

% A_F inside V^perp
% ------------------
%
\draw[dotted, thick, -{Stealth[length=1mm]}] ({projx(0,-1.25)},{projy(0,-1.25,1.75)})
     node[scale=.75] {$\bullet$} --(
     {projx(.8*(-\da*cos(55)+\z),.8*(-\db*sin(55)))/sqrt(2)},
     {projy(.8*(-\da*cos(55)+\z),.8*(-\db*sin(55)),.8*(\da*cos(55)+\z))/sqrt(2)})
     node[above right, scale=.75] {$\hAb_\V$};
\draw ({projx(0,-1.25)},{projy(0,-1.25,1.75)}) node[right, scale=.75] {$\:\: \hAb_\F$};
\end{tikzpicture}
\end{center}
\caption{Schematic representation of the  optimization process.  If $\hAb_\F \in
  \V$, then it is the optimal  estimate $\hAb_\V = \hAb_\F$ (blue situation); If
  $\hAb_\F  \in \V^\star$,  its  closest  point in  $\V$  is $\zerob^\top$  (red
  situation); Otherwise,  $\hAb_\F$ is orthogonally  projected (in the  sense of
  the  $\tFb$-inner  product)  on  the   surface  $\partial\V$  of  $\V$  (black
  situation).}
\label{Fig:ConeConstrainedOptimization}
\end{figure}

Therefore, the surface
\begin{equation*}
\partial\V =  \left\{ (\Ab \,  \big| \, \Ab  \, \Qb_{a,b} \,  \Ab^\top = 0  ) \,
\land \, ( 0 \leqslant \Ab \, \one_z) \right\}
\end{equation*}
is the upper elliptical half-cone with axis $A_z$ and whose vertex is located at
the origin; The  half-cone is a centered half-cone with  a horizontal elliptical
base, with respective semi-axes $\frac{(a-1)  \, z}{a+1}$ and $\frac{\sqrt2 \, b
  \, z}{a+1}$, along the  $x$ and $y$ axes respectively, to  which a rotation of
angle   $\frac{\pi}{4}$  around   the   axis  $A_y$   has   been  applied   (see
Appendix~\ref{App:HalfCone}); $\V$  is the  interior of  this half-cone.   It is
represented in  Figure~\ref{Fig:ConeConstrainedOptimization} for the  case where
the measure of interest is the first invariant.

\

Before going further  in the development of a cone-constrained  detector, it can
be  outlined  that  similar  constraints   were  evidenced  for  square  modulus
measurements  in~\cite{Sheinker2009magnetic}, without  formally identifying  the
cone  equation.  Furthermore, only  white  noise  and  signal projections  on  a
ortho-normalized basis  were considered.   The authors  observed that  the first
coefficient they obtained in the orthogonalized basis was largely dominating, in
the sense it accounted for most of  the signal energy, while the others were not
distinguishable  from noise.   Such a  situation corresponds  to the  case of  a
highly ``slanted''  cone, and the  detector, expressed with the  present paper's
notation reads:
\begin{equation}
\label{Eq:Sheinker}
\hAb_1 =  \begin{bmatrix} \displaystyle\frac{\xb \Fb_1^\top}{\|\Fb_1\|^2} &  0 &
  0 \end{bmatrix}, \qquad
\Lambda_1 = \left\| \xb \, \Fb_1^\top \right\|_F^2 \, \gtrless \, \eta
\end{equation}
where $\Fb_1$ is the vector containing  the discretisation of the first function
in the basis.

This  was shown  to  improve the  detection  performance\footnote{Note that  the
performance is  still given  by expressions~\eqref{Eq:Perf}-\eqref{Eq:StatNaive}
where the degree  of freedom is $1$ instead  of $d \, N$ and  the non centrality
parameter is  $k \:\frac{\left\|  \sb \Fb_1^\top  \right \|_F^2}{\sigma^2  \, \|
  \Fb_1  \|^2}$ instead  of  $k \,  \|  \tsb \|_F^2$,  where  $\sigma^2$ is  the
variance  of  the  noise components.\label{Foot:PerfSheinker}}.   Although  this
result is  of practical interest (in  the white noise case),  a more significant
pitfall arises: unless the projection of $\xb$ onto the basis is zero, we can be
absolutely certain that the estimated signal  has no physical meaning, since $\V
\, \cap \, \left\{ \begin{bmatrix} A_x & 0 & 0 \end{bmatrix} \: \big| \: A_x \in
\Rset \right\}  = \zerob^\top =  \begin{bmatrix} 0 &  0 & 0  \end{bmatrix}$ (the
$x$-axis only intersects the half-cone at the origin, the apex of the cone) and,
as has been  proven previously, all points located outside  the half-cone do not
represent physical signals.

In  a previous  work~\cite{Chenevas2025physics}, we  showed how  to exploit  the
semi-algebraic structure to  construct the receiver in the case  where the noise
is white  Gaussian and where the  coefficients are expressed in  the orthonormal
continuous  basis (orthonormality  assumed to  be  conserved when  the basis  is
sampled  sampled). In  the sequel,  this  approach is  extended to  the case  of
colored noise and discrete basis $\Fb$.

\

The     unconstrained    maximum     likelihood    estimator     was    obtained
in~\eqref{Eq:AHatTilde} as a minimizer of  the estimation error $\left\| \hAb_\F
- \Ab \right\|_{\tFb}^2$, with $\hAb_\F = \txb \, \tFb^\#$, a convex function of
$\Ab$. Since  $\V$ is  also convex,  if a solution  exists in  $\V$, it  will be
unique~\cite{Boyd2004ConvexOptimization}.      Hereafter,    the     constrained
optimization      is      solved      by      the      Lagrange      multipliers
method~\cite{Boyd2004ConvexOptimization,
  Lasserre2015PolynomialOptimization}. Let  $\lambda$, $\mu$ be  the multipliers
associated to the constraints defining the cone $\V$ in Eq.~\eqref{Eq:Cone}; the
optimization problems reads:
\begin{equation*}
\hAb_\V  = \argmin_{\Ab  =  \V}  = \left\|  \hAb_\F  -  \Ab \right\|_{\tFb}^2  =
\argmin_{\Ab \in \Rset^{1 \times 3}} \L\left( \Ab , \lambda , \mu \right)
\end{equation*}
with
\begin{equation*}
\L\left( \Ab , \lambda , \mu \right) = \left\| \hAb_\F - \Ab \right\|_{\tFb}^2 +
\lambda \, \Ab \, \Qb_{a,b} \, \Ab^\top - \mu \, \Ab \, \one_z
\end{equation*}
and where $\lambda,  \: \mu$ are determined to satisfy  the constraints. The two
constraints defining $\V$ are qualified in the Karush-Kuhn-Tucker (KKT) sense if
and  only if\footnote{A  necessary condition  to apply  the KKT  theorem is  the
linear independance of $\pdv*{\Ab \, \Qb_{a,b} \, \Ab^\top}{\Ab}$ and $\pdv*{\Ab
  \,             \one_z}{\Ab}$~\cite[\S~7.1]{Lasserre2015PolynomialOptimization}
or~\cite{Boyd2004ConvexOptimization}.  From a  geometrical point  of view,  this
requires that  the boundary $\partial  \V$ is  differentiable, except at  $\Ab =
\zerob^\top$,   this  latter   case   being  studied   separately.}  $\Ab   \neq
\zerob^\top$.  The gradient of the Lagrangian reads:
\begin{equation*}
\pdv{\L}{\Ab} = 2 \left( \Ab - \hAb_\F  \right) \tFb \, \tFb^\top + 2 \, \lambda
\, \Ab \, \Qb_{a,b} - \mu \, \one_z^\top
\end{equation*}
According      to     the      KKT     theorem~\cite{Boyd2004ConvexOptimization,
  Lasserre2015PolynomialOptimization} the  non zero  extrema of  the constrained
optimization problem satisfy the system below:
\begin{equation*}
\left\{
\begin{array}{l}
\displaystyle \pdv{\L}{\Ab} = 0\\[3mm]
\Ab \,  \Qb_{a,b} \, \Ab^\top  \leqslant 0 \, \land  \, \Ab \,  \one_z \geqslant
0\\[2mm]
\lambda \geq 0 \,  \land \, \mu \geq 0\\[2mm]
\lambda \, \Ab \, \Qb_{a,b} \, \Ab^\top = 0 \, \land \, \mu \, \Ab \, \one_z = 0
\end{array}
\right.
\raisetag{2\normalbaselineskip}
\end{equation*}
Thus, $\mu \ne  0$ (the second constraint  is saturated) implies $A_z  = 0$; the
first constraint $\Ab \, \Qb_{a,b} \, \Ab^\top \leqslant 0$ now reads $2 \, a \,
b^2 A_x^2 + 2 \left(  a^2 - 1 \right) A_y^2 \leqslant 0$.  Since $a  > 1, \: b >
0$,  necessarily $A_x  =  A_y =  0$, which  contradicts  $\Ab \ne  \zerob^\top$.
Finally $\mu = 0$, thus leading to the simplified system:
\begin{equation*}
\left\{
\begin{array}{l}
\Ab - \hAb_\F + \lambda \Ab \, \Qb_{a,b} \left( \tFb \, \tFb^\top \right)^{-1} =
0\\[3mm]
\Ab \,  \Qb_{a,b} \, \Ab^\top  \leqslant 0 \, \land  \, \Ab \,  \one_z \geqslant
0\\[2mm]
\lambda \geq 0\\[2mm]
\lambda \, \Ab  \, \Qb_{a,b} \, \Ab^\top =  0
\end{array}
\right.
\raisetag{2\normalbaselineskip}
\end{equation*}
%
% Q* = R diag [2 (a-1)^2  &  b^2   &  2 (a+1)^2 ] R^t
%
Two  cases  arise when  considering  the  saturation/unsaturation of  the  first
constraint.
\begin{itemize}
  \item Case  $\lambda = 0$: $\hAb_\V  = \hAb_\F$ is obtained;  this solution is
    admissible if and only if $\hAb_\F \in \V$.
  \item case  $\lambda \ne 0$  (the constraint is  saturated); the system  to be
    solved is:
    \begin{equation}
    \label{Eq:LagrangianSystem}
    \left\{
    \begin{array}{l}
    \Ab  - \hAb_\F  +  \lambda \,  \Ab  \, \Qb_{a,b}  \left(  \tFb \,  \tFb^\top
    \right)^{-1} = 0\\[3mm]
    \Ab \,  \Qb_{a,b} \, \Ab^\top  = 0 \, \land  \, \Ab \,  \one_z \geqslant
    0\\[2mm]
    \lambda > 0
    \end{array}
    \right.
    \raisetag{2\normalbaselineskip}
    \end{equation}
    For $\Ab \ne 0$, the equality  $\Ab \Qb_{a,b} \Ab^\top = 0$ (expressing that
    $\Ab \in \partial \V$) can be reformulated as $\left\langle \Ab \, \Qb_{a,b}
    \left( \tFb  \, \tFb^\top \right)^{-1}  , \: \Ab \right\rangle_{\tFb}  = 0$.
    Consequently, the first  equation of the system  above imposes $\left\langle
    \Ab - \hAb_\F \: , \: \Ab \right\rangle_{\tFb} = 0$. This equality, together
    with the  constraint $\Ab \,  \one_z \geqslant 0$  can be satisfied  only if
    $\hAb_\F  \in  \partial\V_\Fb^\perp$,  where $\partial\V_\Fb^\perp$  is  the
    half-space  (as  $\Ab  \,  \one_z   \geqslant  0$)  containing  all  vectors
    orthogonal (w.r.t.  $\tFb$-inner product) to $\partial\V$. Thus:
    \begin{itemize}
    \item If  $\hAb_\F \in  \partial\V_\Fb^\perp$ and $\lambda  > 0$:  The first
      equality of the system~\eqref{Eq:LagrangianSystem} can be reformulated as:
      \begin{equation}
      \label{Eq:HA_Vperp}
      \hAb_\V = \hAb_\F \, \left( \Ib_3 + \lambda_\V \, \Qb_{a,b} \left( \tFb \,
      \tFb^\top \right)^{-1} \right)^{-1}
      \end{equation}
      where $\lambda_\V$  is determined to satisfy  the constraint\footnote{Note
      that the uniqueness of the solution prevents $-\lambda_\V^{-1}$ from being
      an  eigenvalue of  $\Qb_{a,b}  \left( \tFb  \, \tFb^\top  \right)^{-1}$.}.
      Subsequently, by substituting the equality above in second equation of the
      system~\eqref{Eq:LagrangianSystem},       we       show       in       the
      appendix~\ref{App:QuarticPolynomial}  that  $\lambda_\V$   is  root  of  a
      quartic polynomial.  The  roots of this polynomial are  obtained thanks to
      Ferrari's  method~\cite[\S~3.8.3]{Abramowitz1972handbook},  and  only  the
      (necessarily unique) positive root is retained for $\lambda_\V$.
    \item If  conversely $\hAb_\F  \in \V^\star =  \Rset^{1 \times  3} \setminus
      \left(  \V  \cup \partial\V^\perp_\Fb  \right)$  and  $\lambda >  0$:  the
      constraints are not qualified (in the  KKT sense) and therefore $\hAb_\V =
      \zerob^\top$, which corresponds to the physically admissible solution that
      is the closest (w.r.t. $\tFb$-norm) to $\Ab_\F$.
    \end{itemize}
\end{itemize}

To  summarize,  estimating  $\hAb_\V$  can   be  interpreted  from  a  geometric
perspective as illustrated in Fig.~\ref{Fig:ConeConstrainedOptimization}.  After
the unconstrained maximum likelihood estimator (UMLE) $\hAb_\F$ is computed, one
among  the three  situations below  may arise  and the  estimation algorithm  is
sketched below:

\begin{algorithm}
\caption{$\hAb_\V$ geometric estimation}
\begin{algorithmic}
\State Compute UMLE $\hAb_\F$\vspace{1mm}
\If{\: $\hAb_\F \in \V$ \:}\vspace{1mm}
\State $\hAb_\V =\hAb_\F$ \vspace{1mm}
\ElsIf{\: $\hAb_\F  \in \V^\star$ \:}\vspace{1mm}
\State $\hAb_\V = \zerob^\top$\vspace{1mm}
\Else\vspace{1mm}%{$\hAb_\F  \not\in \V  \cup  \V^\star$}
\State  {$\hAb_\V$  is  the   orthogonal  projection\footnote{In  the  sense  of
  $\tFb$-inner  product,  which  requires  to  solve  the  quartic  equation  in
  $\lambda$.} onto $\partial\V$.}\vspace{1mm}
\EndIf   
\end{algorithmic}
\end{algorithm}

{\em  Remark}: $\V^\star$  is  the interior  of  the surface  $\partial\V^\star$
determined  by the  set of  vectors  $\Ab$ orthogonal  to $\Ab_{\partial\V}  \in
\partial \V$. Thus  $\Ab \left( \Fb \, \Fb^\top  \right) \Ab_{\partial\V}^\top =
0$. Since $\left\langle \Ab_{\partial \V}  \, \Qb_{a,b} \left( \tFb \, \tFb^\top
\right)^{-1}, \Ab{\partial \V} \right\rangle_{\tFb} =  0$, we conclude that $\Ab
\left( \Fb \, \Fb^\top \right)$  and $ \Ab_{\partial\V} \Qb_{a,b}$ are collinear
vectors,  or equivalently  $\Ab \left(  \Fb \,  \Fb^\top \right)  \Qb_{a,b}^{-1}
\propto  \Ab_{\partial\V}$.   By  definition, $  \Ab_{\partial\V}  \,  \Qb_{a,b}
\Ab_{\partial\V}^\top = 0$, and $\V^\star$ is characterized by
\begin{equation*}
\V^\star =  \left\{ \left. \Ab  = \Bb \left(  \tFb \, \tFb^\top  \right)^{-1} \:
\right| \: \Bb \in \V^\star_{\Qb} \right\}
\end{equation*}
with
\begin{equation*}
\V^\star_{\Qb}  =  \left\{  \Bb \in  \Rset^{1  \times  3}  \:  \big| \:  \Bb  \,
\Qb^{-1}_{a,b} \,  \Bb^\top \geqslant  0 \,  \land \,  \Bb \,  \Qb_{a,b}^{-1} \,
\one_z \leqslant 0 \right\}
\end{equation*}

%% DEJA DIT {\em Remark}: An alternate strategy can be considered, consisting in
%% finding  the values  that  make  the gradient  (with  respect  to the  dipole
%% moments)  vanish. This  approach requires  solving a  multivariate polynomial
%% system with  three unknowns, of  maximum degree 3, which  entails significant
%% computational  difficulties. The  choice of  the Lagrangian  method developed
%% earlier offers  two major  advantages. First,  it allows  us to  identify the
%% exact  maximizer; second,  the  analytical solution  of  the problem  enables
%% extremely fast processing, paving the way for real-time processing.

% --------------------------- Simulation results ----------------------------- %

\section{Simulation results}
\label{Sec:Simulation}

In this section, we propose to evaluate the performance of the GLRT $\Lambda_\V$
(constrained estimation  of $\Ab$, Eq.~\eqref{Eq:GLLRT-V}). It  will be compared
to the naive receiver $\Lambda_\F$ (using an unconstrained parameter estimation,
Eq.~\eqref{Eq:GLLRT-Naive}),  and  to   the  clear-seeing  receiver  $\Lambda_c$
(completely known  signal, Eq.~\eqref{Eq:LLRT})  as a  reference that  cannot be
overpassed. For the measurement of the square modulus of the anomaly and for the
first invariant of  the gradiant tensor, the results obtained  are compared with
those obtained by Sheinker's method~\cite{Sheinker2009magnetic} ($\Lambda_1$, in
Eq.~\eqref{Eq:Sheinker}.)

All  simulations  below rely  on  randomly  selected  dipole sources  $\mb$  and
trajectories $(\alpha, \beta)$. Typical simulation parameter values are reported
in  Table~\ref{Tab:SimulationParameters}; they  are tuned  in order  to mimic  a
realistic airborne  MAD framework. Furthermore,  without loss of  generality, we
consider  a  centered  Gaussian  white noise  with  diagonal  covariance  matrix
$\sigma^2  \Ib$ (remind  that it  was  outlined in  a preceding  section that  a
whitening step may  be applied).  This allows a fair  comparison of the proposed
approach with Sheinker's method using  $\Lambda_1$, since this latter assumes an
observation model with additive Gaussian white noise.
\begin{table}[!h]
\begin{center}
\begin{tabular}{|c|c|}
\hline                                   $V$                                   &
$92.5~\si[per-mode=reciprocal,inter-unit-product=.]{\meter\per\second}$ \\
\hline $D$ & $412.32~\si{\meter}$  \\
\hline $K$ & $1000$ samples \\
\hline $f_s$ & $10~\si{\hertz}$ \\
\hline $\alpha$ & $2.77~\si{\radian}$ \\
\hline $\beta$ & $0.52~\si{\radian}$ \\
\hline             $\mb$            &             $[-0.25,\,            -1.02,\,
  -0.55]^\top~\si[per-mode=reciprocal,inter-unit-product=.]{\ampere.\meter^2}$\\
\hline
\end{tabular}
\end{center}
\caption{Simulation parameters}
\label{Tab:SimulationParameters}
\end{table}
The signal-to-noise  ratio (SNR), used  to tune  the simulation noise  level, is
defined by
\begin{equation*}
\SNR (\dB) = 10 \log_{10}  \left( \dfrac{\left\| \tsb \right\|_F^2}{d K} \right)
, \qquad \tsb = \sigma^{-1} \sb
\end{equation*}

\begin{figure}[ht]
\input{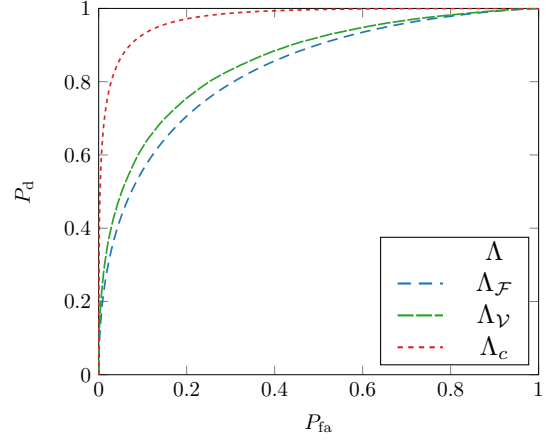}
\caption{Vector  measurement of  the  anomaly:  ROC with  $\SNR  = -26~\dB$  for
  receivers,   estimated  from   Monte  Carlo   simulation  with   $10^5$  noise
  realizations}
\label{Fig:ROC-3D}
\end{figure}

\begin{figure}[ht]
\input{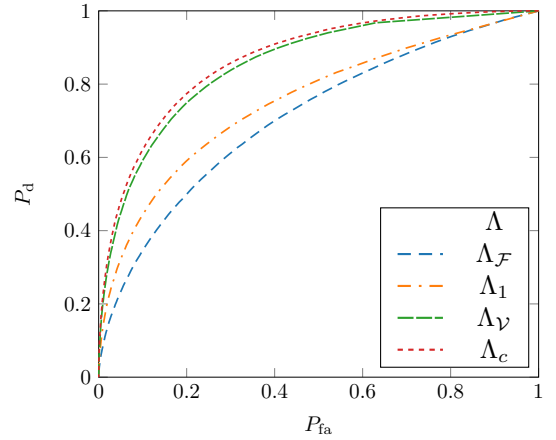}
\caption{Square module of the anomaly: ROC  with $\SNR = -26~\dB$ for receivers,
  estimated from Monte Carlo simulation with $10^5$ noise realizations}
\label{Fig:ROC-Square}
\end{figure}

\begin{figure}[ht]
\input{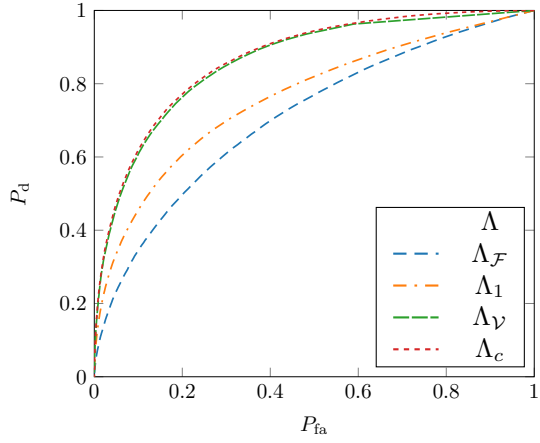}
\caption{First principal  invariant of the  magnetic gradient: ROC with  $\SNR =
  -26~\dB$  for receivers,  estimated from  Monte Carlo  simulation with  $10^5$
  noise realizations}
\label{Fig:ROC-I1}
\end{figure}

The theoretical performances for $\Lambda_c$, $ \Lambda_\F$, and $\Lambda_1$ are
obtained       directly      by       applying       the      results       from
sub-subsection~\ref{SubSubSec:Performance}               (see               also
footnote~\ref{Foot:PerfSheinker}  for  $\Lambda_1$).   However,  the  analytical
derivation of  the performances of $\Lambda_\V$  do not enter the  framework and
assumptions  from   section~\ref{SubSubSec:Performance}  and   are  subsequently
obtained  by Monte-Carlo  simulation  involving $10^5$  noise realizations;  the
receiver operating characteristic  (ROC) curves for a  representative source are
reported in figures~\ref{Fig:ROC-3D},~\ref{Fig:ROC-Square} and~\ref{Fig:ROC-I1},
corresponding to the $3-$dimensional magnetometer  case with $\Pib = \Ib_3$, the
square modulus case and the first invariant case, respectively.

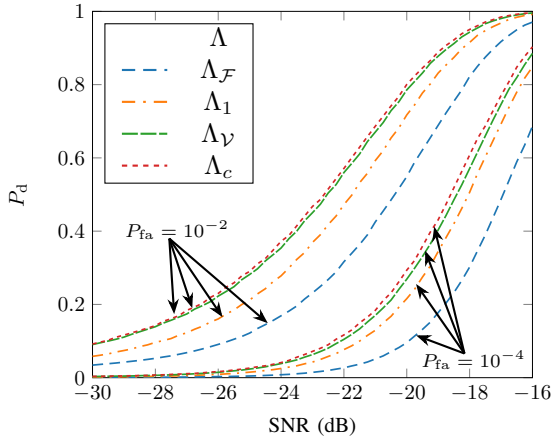
\begin{figure}[ht]
\begin{center}
    \begin{tikzpicture}[scale=.8]
  \begin{axis}[
  name=boundary,
  width=\columnwidth,
    xlabel={$\SNR$ (dB)},
    ylabel={$\Pd$},
    xmin=-30, xmax=-16,
    ymin=0, ymax=1,
    xtick={-30, -28, -26, -24, -22, -20, -18, -16},
    ytick={0,0.20,0.40,0.60,0.80,1.0},
  ]

%%%%%%%%%%%%%%%%%%%%%%%%%%%%%%%%%%%%%%%%%%%%%%%%%%%%%%
% % pfa = 0.001

\addplot[
    color=bleu,
    dash pattern=on 5pt off 3pt,
    thick
]
coordinates {
(-30.0,0.00054) (-29.75,0.00055) (-29.5,0.00062) (-29.25,0.00074) (-29.0,0.00079) (-28.75,0.00089) (-28.5,0.00097) (-28.25,0.00107) (-28.0,0.00112) (-27.75,0.0013) (-27.5,0.00141) (-27.25,0.00159) (-27.0,0.00178) (-26.75,0.00193) (-26.5,0.00219) (-26.25,0.00261) (-26.0,0.00292) (-25.75,0.0032) (-25.5,0.00363) (-25.25,0.00398) (-25.0,0.00459) (-24.75,0.00521) (-24.5,0.00606) (-24.25,0.00692) (-24.0,0.00801) (-23.75,0.00905) (-23.5,0.01067) (-23.25,0.01227) (-23.0,0.01455) (-22.75,0.01695) (-22.5,0.0197) (-22.25,0.02332) (-22.0,0.02685) (-21.75,0.03205) (-21.5,0.0368) (-21.25,0.04282) (-21.0,0.05116) (-20.75,0.05945) (-20.5,0.07088) (-20.25,0.08226) (-20.0,0.09559) (-19.75,0.11361) (-19.5,0.13176) (-19.25,0.15167) (-19.0,0.17766) (-18.75,0.20378) (-18.5,0.23223) (-18.25,0.26529) (-18.0,0.30199) (-17.75,0.34619) (-17.5,0.38969) (-17.25,0.43799) (-17.0,0.48696) (-16.75,0.53769) (-16.5,0.58916) (-16.25,0.63461) (-16.0,0.68655)
};
\label{pgfplots2:c1}    
    
\addplot[
    color=orange,
    dash pattern=on 5pt off 3pt on 1pt off 3pt,
    thick
]
coordinates {
(-30.0,0.00179) (-29.75,0.00195) (-29.5,0.0022) (-29.25,0.00232) (-29.0,0.00254) (-28.75,0.00281) (-28.5,0.00306) (-28.25,0.00346) (-28.0,0.00374) (-27.75,0.00408) (-27.5,0.00458) (-27.25,0.00511) (-27.0,0.00592) (-26.75,0.00644) (-26.5,0.00722) (-26.25,0.00825) (-26.0,0.00914) (-25.75,0.01032) (-25.5,0.01152) (-25.25,0.01315) (-25.0,0.01507) (-24.75,0.01705) (-24.5,0.01963) (-24.25,0.02193) (-24.0,0.02512) (-23.75,0.02888) (-23.5,0.03239) (-23.25,0.03788) (-23.0,0.0426) (-22.75,0.0497) (-22.5,0.05795) (-22.25,0.06582) (-22.0,0.07665) (-21.75,0.08703) (-21.5,0.09996) (-21.25,0.11537) (-21.0,0.12942) (-20.75,0.1495) (-20.5,0.16756) (-20.25,0.19251) (-20.0,0.21569) (-19.75,0.24155) (-19.5,0.27466) (-19.25,0.30496) (-19.0,0.34455) (-18.75,0.38253) (-18.5,0.42039) (-18.25,0.46195) (-18.0,0.50587) (-17.75,0.55738) (-17.5,0.60284) (-17.25,0.64796) (-17.0,0.69278) (-16.75,0.73765) (-16.5,0.77946) (-16.25,0.81754) (-16.0,0.84841)
};
\label{pgfplots2:c2}

\addplot[
    color=vert,
    dash pattern=on 6pt off 1pt,
    thick
]
coordinates {
(-30.0,0.00308) (-29.75,0.00351) (-29.5,0.0038) (-29.25,0.00415) (-29.0,0.0044) (-28.75,0.00482) (-28.5,0.00541) (-28.25,0.00592) (-28.0,0.00634) (-27.75,0.00698) (-27.5,0.0077) (-27.25,0.00857) (-27.0,0.00948) (-26.75,0.01063) (-26.5,0.01177) (-26.25,0.0132) (-26.0,0.01456) (-25.75,0.01687) (-25.5,0.0187) (-25.25,0.02071) (-25.0,0.02359) (-24.75,0.0264) (-24.5,0.02969) (-24.25,0.03414) (-24.0,0.03849) (-23.75,0.04326) (-23.5,0.04853) (-23.25,0.05612) (-23.0,0.06297) (-22.75,0.07136) (-22.5,0.0806) (-22.25,0.09294) (-22.0,0.10465) (-21.75,0.11817) (-21.5,0.13321) (-21.25,0.15202) (-21.0,0.17014) (-20.75,0.19049) (-20.5,0.21184) (-20.25,0.24124) (-20.0,0.2689) (-19.75,0.29897) (-19.5,0.33072) (-19.25,0.36502) (-19.0,0.40339) (-18.75,0.44296) (-18.5,0.48545) (-18.25,0.52635) (-18.0,0.57086) (-17.75,0.6155) (-17.5,0.66024) (-17.25,0.70419) (-17.0,0.74612) (-16.75,0.78736) (-16.5,0.81922) (-16.25,0.85428) (-16.0,0.88515)
};
\label{pgfplots2:c3}
    
\addplot[
    color=rouge,
    dash pattern=on 2pt off 2pt, 
    thick
]
coordinates {
(-30.0,0.00354) (-29.75,0.00382) (-29.5,0.00407) (-29.25,0.00464) (-29.0,0.00494) (-28.75,0.00528) (-28.5,0.00589) (-28.25,0.00629) (-28.0,0.00731) (-27.75,0.00785) (-27.5,0.00894) (-27.25,0.00972) (-27.0,0.01092) (-26.75,0.01164) (-26.5,0.01317) (-26.25,0.01523) (-26.0,0.01645) (-25.75,0.01844) (-25.5,0.0209) (-25.25,0.02329) (-25.0,0.02619) (-24.75,0.02978) (-24.5,0.03243) (-24.25,0.03661) (-24.0,0.04119) (-23.75,0.04713) (-23.5,0.05581) (-23.25,0.06329) (-23.0,0.07175) (-22.75,0.08018) (-22.5,0.08988) (-22.25,0.10488) (-22.0,0.11673) (-21.75,0.13036) (-21.5,0.14959) (-21.25,0.16547) (-21.0,0.18386) (-20.75,0.21062) (-20.5,0.23166) (-20.25,0.26191) (-20.0,0.29349) (-19.75,0.31997) (-19.5,0.35871) (-19.25,0.39756) (-19.0,0.43113) (-18.75,0.47369) (-18.5,0.51852) (-18.25,0.56292) (-18.0,0.60773) (-17.75,0.65269) (-17.5,0.68737) (-17.25,0.72951) (-17.0,0.7702) (-16.75,0.80773) (-16.5,0.84754) (-16.25,0.87689) (-16.0,0.90306)
};
\label{pgfplots2:c4}

%%%%%%%%%%%%%%%%%%%%%%%%%%%%%%%%%%%%%%%%%%%%
% pfa = 0.1

\addplot[
    color=bleu,
    dash pattern=on 5pt off 3pt,
    thick
]
coordinates {
(-30.0,0.03428) (-29.75,0.03627) (-29.5,0.03815) (-29.25,0.04008) (-29.0,0.04243) (-28.75,0.04478) (-28.5,0.04716) (-28.25,0.04989) (-28.0,0.05279) (-27.75,0.05638) (-27.5,0.05963) (-27.25,0.06388) (-27.0,0.0683) (-26.75,0.07316) (-26.5,0.07852) (-26.25,0.08452) (-26.0,0.09083) (-25.75,0.0974) (-25.5,0.10494) (-25.25,0.1125) (-25.0,0.12173) (-24.75,0.13422) (-24.5,0.14481) (-24.25,0.15635) (-24.0,0.16898) (-23.75,0.18167) (-23.5,0.19571) (-23.25,0.21473) (-23.0,0.23142) (-22.75,0.24901) (-22.5,0.26796) (-22.25,0.28866) (-22.0,0.31644) (-21.75,0.3397) (-21.5,0.36359) (-21.25,0.39016) (-21.0,0.42536) (-20.75,0.45492) (-20.5,0.48529) (-20.25,0.5167) (-20.0,0.54892) (-19.75,0.58829) (-19.5,0.62173) (-19.25,0.65522) (-19.0,0.68916) (-18.75,0.72275) (-18.5,0.75456) (-18.25,0.78543) (-18.0,0.82212) (-17.75,0.84989) (-17.5,0.875) (-17.25,0.89729) (-17.0,0.9174) (-16.75,0.93428) (-16.5,0.94957) (-16.25,0.96228) (-16.0,0.97066)
};
\label{pgfplots2:c1}  

\addplot[
    color=orange,
    dash pattern=on 5pt off 3pt on 1pt off 3pt,
    thick
]
coordinates {
(-30.0,0.05824) (-29.75,0.06135) (-29.5,0.06488) (-29.25,0.07012) (-29.0,0.07416) (-28.75,0.07844) (-28.5,0.08478) (-28.25,0.08951) (-28.0,0.09642) (-27.75,0.10129) (-27.5,0.10682) (-27.25,0.11609) (-27.0,0.12253) (-26.75,0.13231) (-26.5,0.14049) (-26.25,0.15178) (-26.0,0.16035) (-25.75,0.17354) (-25.5,0.18476) (-25.25,0.19975) (-25.0,0.2119) (-24.75,0.22878) (-24.5,0.24281) (-24.25,0.26231) (-24.0,0.27796) (-23.75,0.29965) (-23.5,0.32287) (-23.25,0.34162) (-23.0,0.36922) (-22.75,0.39017) (-22.5,0.41797) (-22.25,0.44121) (-22.0,0.47256) (-21.75,0.49826) (-21.5,0.53233) (-21.25,0.5582) (-21.0,0.59288) (-20.75,0.61966) (-20.5,0.65437) (-20.25,0.68187) (-20.0,0.71734) (-19.75,0.7448) (-19.5,0.77731) (-19.25,0.80199) (-19.0,0.83169) (-18.75,0.85354) (-18.5,0.87432) (-18.25,0.89789) (-18.0,0.91509) (-17.75,0.93456) (-17.5,0.94745) (-17.25,0.95855) (-17.0,0.96819) (-16.75,0.97787) (-16.5,0.98386) (-16.25,0.98865) (-16.0,0.99197)
};
\label{pgfplots2:c2}

\addplot[
    color=vert,
    dash pattern=on 6pt off 1pt,
    thick
]
coordinates {
(-30.0,0.09039) (-29.75,0.0953) (-29.5,0.09999) (-29.25,0.10496) (-29.0,0.11073) (-28.75,0.11854) (-28.5,0.12537) (-28.25,0.13228) (-28.0,0.13915) (-27.75,0.14893) (-27.5,0.15676) (-27.25,0.1654) (-27.0,0.17398) (-26.75,0.18726) (-26.5,0.1967) (-26.25,0.20736) (-26.0,0.22265) (-25.75,0.23521) (-25.5,0.24817) (-25.25,0.2675) (-25.0,0.28176) (-24.75,0.29718) (-24.5,0.31869) (-24.25,0.33592) (-24.0,0.35398) (-23.75,0.37939) (-23.5,0.40004) (-23.25,0.42837) (-23.0,0.4504) (-22.75,0.47339) (-22.5,0.50422) (-22.25,0.52707) (-22.0,0.55994) (-21.75,0.58487) (-21.5,0.61073) (-21.25,0.6447) (-21.0,0.67059) (-20.75,0.69655) (-20.5,0.72895) (-20.25,0.7534) (-20.0,0.78557) (-19.75,0.80866) (-19.5,0.83177) (-19.25,0.85816) (-19.0,0.8777) (-18.75,0.89597) (-18.5,0.91206) (-18.25,0.93022) (-18.0,0.94319) (-17.75,0.95465) (-17.5,0.96639) (-17.25,0.9741) (-17.0,0.98054) (-16.75,0.98605) (-16.5,0.99011) (-16.25,0.99356) (-16.0,0.99573)
};
\label{pgfplots2:c3}
    
\addplot[
    color=rouge,
    dash pattern=on 2pt off 2pt, 
    thick
]
coordinates {
(-30.0,0.0919) (-29.75,0.09706) (-29.5,0.10228) (-29.25,0.11051) (-29.0,0.11627) (-28.75,0.12237) (-28.5,0.12889) (-28.25,0.13525) (-28.0,0.14199) (-27.75,0.15387) (-27.5,0.16095) (-27.25,0.16936) (-27.0,0.18368) (-26.75,0.19299) (-26.5,0.20305) (-26.25,0.21917) (-26.0,0.22991) (-25.75,0.24711) (-25.5,0.25876) (-25.25,0.27788) (-25.0,0.29005) (-24.75,0.31085) (-24.5,0.32499) (-24.25,0.34842) (-24.0,0.37229) (-23.75,0.38821) (-23.5,0.4146) (-23.25,0.44124) (-23.0,0.46748) (-22.75,0.48629) (-22.5,0.51492) (-22.25,0.54324) (-22.0,0.57114) (-21.75,0.60069) (-21.5,0.62924) (-21.25,0.65833) (-21.0,0.68664) (-20.75,0.71394) (-20.5,0.7419) (-20.25,0.76819) (-20.0,0.80051) (-19.75,0.82369) (-19.5,0.84546) (-19.25,0.86573) (-19.0,0.88912) (-18.75,0.90693) (-18.5,0.92199) (-18.25,0.93889) (-18.0,0.95054) (-17.75,0.96075) (-17.5,0.97084) (-17.25,0.97747) (-17.0,0.98457) (-16.75,0.98853) (-16.5,0.99208) (-16.25,0.99456) (-16.0,0.99641)
};
\label{pgfplots2:c4}

\end{axis}

 \node[draw,fill=white,inner sep=0pt,above left=0.5em, anchor=north west] at ([xshift=10pt,yshift=-10pt]boundary.north west) {
    \begin{tabular}{ccc}
 & $\Lambda$ \\
    \ref{pgfplots2:c1} & $\Lambda_\F$\\
    \ref{pgfplots2:c2} & $\Lambda_1$ \\
    \ref{pgfplots2:c3} & $\Lambda_\V$ \\
    \ref{pgfplots2:c4} & $\Lambda_c$
    \end{tabular}
    };

\draw[-{Stealth[length=2mm]}, thick, black] (6.15,0.4) -- (5.35,0.70);   
\draw[-{Stealth[length=2mm]}, thick, black] (6.15,0.4) -- (5.35,1.55);
\draw[-{Stealth[length=2mm]}, thick, black] (6.15,0.4) -- (5.5,2.10);
\draw[-{Stealth[length=2mm]}, thick, black] (6.15,0.4) -- (5.65,2.50);
\node[black] at (6.3,0.3) {\scriptsize $\Pfa=10^{-4}$};

\node[black] at (1.4,2.45) {\scriptsize $\Pfa=10^{-2}$}; 
\draw[-{Stealth[length=2mm]}, thick, black] (1.27,2.30) -- (1.35,1.05);   
\draw[-{Stealth[length=2mm]}, thick, black] (1.27,2.30) -- (1.65,1.15);
\draw[-{Stealth[length=2mm]}, thick, black] (1.27,2.30) -- (2.15,1.00);
\draw[-{Stealth[length=2mm]}, thick, black] (1.27,2.30) -- (2.90,0.90);
    
\end{tikzpicture}
\end{center}
\caption{Square modulus of  the anomaly measurement: detection  performance as a
function of  SNR with  $\Pfa =  10^{-2}, \,  10^{-4}$ calculated  by Monte-Carlo
simulation with $10^5$ noise realizations.}
%The  shapes of  the curves  are  identical for  the vector  measurement of  the
%anomaly and first principal invariant of the magnetic gradient.}
\label{Fig:Pd-Square-SNR}
\end{figure}

These curves clearly highlight that the  proposed approach improves on all other
approches.  Moreover in  Figures ~\ref{Fig:ROC-Square} and~\ref{Fig:ROC-I1}, the
performance is  extremely close  to the clear-seeing  receiver, which  given the
nature of the latter, is a major advance.  This statement remains valid when the
signal-to-noise      ratio     (SNR)      varies,     as      illustrated     in
Figure~\ref{Fig:Pd-Square-SNR},   which   shows   the   results   obtained   for
measurements of the square modulus.

To avoid  overloading the figures  with identical content, the  analogous curves
for the case of  the vector anomaly and the measurement  of the tensor invariant
are not presented here, but they  exhibit similar behavior. Their shapes and the
conclusions that  can be drawn  for these  two other measurement  strategies are
identical to those obtained in the case of the square modulus measurement of the
anomaly.

These  results  allow  us  to   emphasize  the  importance  of  taking  physical
constraints (namely the  geometric structure of the  space containing physically
meaning-full dipole signals) into account  when designing a better detector: the
identification of the (slanted) cone  of physically admissible dipole parameters
and the  subsequent estimation algorithm seem  to be of paramount  importance in
the detector design.

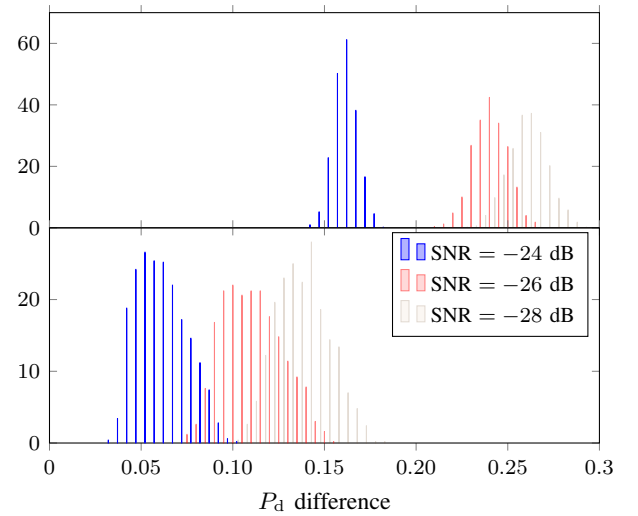
\begin{figure}[ht]
\begin{center}
    \begin{tikzpicture}
    \begin{groupplot}[
        ybar,
        group style={
            group name=auc,
            group size=1 by 2,
            xlabels at=edge bottom,
            xticklabels at=edge bottom,
            x descriptions at=edge bottom,
            y descriptions at=edge left,
            vertical sep=0.0cm,
        },
        footnotesize,
        width=\columnwidth,
        height=0.5\columnwidth,
        xlabel=$\Pd$ difference,
       xtick={0, 0.05, 0.10, 0.15, 0.20, 0.25, 0.3},
       xticklabels={$0$, $0.05$, $0.10$, $0.15$, $0.20$, $0.25$, $0.3$},
       xmax=0.3,
       xmin=0,
       ymin=0,
    ]
    
\nextgroupplot[ymax=70,ytick={0, 20, 40, 60},]
\addplot +[bar width=0.001,opacity=1] table {
0.14500000000000005 0.9999999999999991
0.15000000000000005 5.199999999999996
0.15500000000000005 22.79999999999998
0.16000000000000006 50.19999999999995
0.16500000000000006 61.199999999999946
0.17000000000000007 38.19999999999996
0.17500000000000007 16.599999999999987
0.18000000000000008 4.599999999999996
0.18500000000000008 0.19999999999999982
};
\addplot +[bar width=0.001,opacity=0.5] table {
0.2100000000000001 0.39999999999999963
0.2150000000000001 1.1999999999999988
0.2200000000000001 4.799999999999995
0.22500000000000012 9.999999999999991
0.23000000000000012 26.799999999999976
0.23500000000000013 34.99999999999997
0.24000000000000013 42.39999999999996
0.24500000000000013 34.00000000000016
0.2500000000000001 26.399999999999977
0.2550000000000001 13.199999999999989
0.2600000000000001 3.9999999999999964
0.2650000000000001 1.7999999999999985
};
\addplot +[bar width=0.001,opacity=0.2] table {
0.22500000000000012 0.19999999999999982
0.23000000000000012 0.5999999999999994
0.23500000000000013 3.9999999999999964
0.24000000000000013 9.79999999999999
0.24500000000000013 17.20000000000008
0.2500000000000001 25.79999999999998
0.2550000000000001 36.59999999999997
0.2600000000000001 37.19999999999996
0.2650000000000001 30.99999999999997
0.27000000000000013 20.19999999999998
0.27500000000000013 9.59999999999999
0.28000000000000014 5.7999999999999945
0.28500000000000014 1.7999999999999985
0.29000000000000015 0.19999999999999982
};

\nextgroupplot[ymax=30, ytick={0, 10, 20},]
\addplot +[bar width=0.001,opacity=1] table {
0.035 0.40000000000000024
0.04 3.4000000000000017
0.045 18.799999999999983
0.05 24.20000000000001
0.055 26.600000000000016
0.06 25.399999999999977
0.065 25.199999999999978
0.07 22.00000000000004
0.075 17.199999999999985
0.08 14.599999999999987
0.085 11.200000000000022
0.09 7.399999999999993
0.095 2.7999999999999976
0.1 0.6000000000000011
0.105 0.19999999999999982
};
\addlegendentry{$\SNR=-24$ dB};
\addplot +[bar width=0.001,opacity=0.5] table {
0.075 1.1999999999999988
0.08 2.599999999999998
0.085 7.600000000000015
0.09 16.799999999999986
0.095 21.19999999999998
0.1 22.00000000000004
0.105 20.59999999999998
0.11 21.19999999999998
0.115 21.20000000000004
0.12 17.599999999999987
0.125 14.799999999999986
0.13 11.39999999999999
0.135 9.199999999999992
0.14 7.800000000000036
0.145 2.9999999999999973
0.15 1.5999999999999985
0.155 0.19999999999999982
};
\addlegendentry{$\SNR=-26$ dB};
\addplot +[bar width=0.001,opacity=0.2] table {
0.1 0.40000000000000074
0.105 2.599999999999998
0.11 5.7999999999999945
0.115 12.200000000000024
0.12 19.59999999999998
0.125 22.99999999999998
0.13 24.99999999999998
0.135 22.39999999999998
0.14 28.00000000000013
0.145 18.59999999999998
0.15 14.399999999999988
0.155 13.399999999999988
0.16 6.999999999999994
0.165 4.799999999999995
0.17 2.3999999999999977
0.17500000000000002 0.20000000000000093
0.18 0.19999999999999982
};
\addlegendentry{$\SNR=-28$ dB};

\end{groupplot}

\end{tikzpicture}
\end{center}
\caption{Square modulus  of the  anomaly: normalized  distribution of  the $\Pd$
  difference between our approach~\eqref{Eq:GLLRT-V}-\eqref{Eq:Ah} and classical
  one~\eqref{Eq:GLLRT-Naive}      (top       panel),      dominant      function
  one~\cite{Sheinker2009magnetic}  (bottom  panel),  for different  SNR;  $10^3$
  randomly chosen  dipole moments  are used for  the Monte-Carlo  simulation. In
  each case, the $\Pd$'s were calculated for $10^4$ noise realizations.}
\label{Fig:Pd-diff}
\end{figure}

Since we have examined only a  single source and trajectory configuration above,
it  is necessary  to consider  a much  broader set  of simulations  in order  to
confirm and  illustrate the  performance of our  algorithm.  In  the simulations
shown in  Figure~\ref{Fig:Pd-diff}, square modulus measurements  are considered;
we set $\Pfa = 10^{-2}$ and evaluate the relative gain in terms of the detection
probability        between       our        algorithm       and        Sheiker's
algorithm~\cite{Sheinker2009magnetic} (bottom  panel) between our  algorithm and
the  classical  unconstrained GLRT  algorithm  (top  panel). A  thousand  dipole
moments was  randomly selected, and  the probability of detection  was estimated
using a Monte-Carlo simulation involving $10^4$ noise realizations. The interest
of  our  approach appears  clearly,  as  the  difference between  the  detection
probabilities achieved by our method and the other method is always positive and
significant.   Furthermore, the  advantage  of  physically constraint  detection
appears to increase  when the SNR worsen, which is  expected: actually, the mean
squared error of the estimated parameters increases for low SNR, possibly giving
even non physical solutions when this  constraint is ignored.  The histograms of
the  differences in  $\Pb$  allow  us to  conclude  that  the preceding  results
presented for a single source example are indeed generic.

{\em Remarks}:
\begin{itemize}
\item  The  same simulations  completed  with  the  clear-seeing receiver  as  a
  reference, confirms that this latter and our approach give very close results,
  since the differences in $\Pd$  remain between $0$ and $5.10^{-2}$, regardless
  of the SNR.
  % which represents a very small deviation.
  %
\item When  using a vector measure  of the anomaly, the  results obtained appear
  less impressive.  In fact, for certain implementations, the classical approach
  proves to be more effective than  the constrained approach. This problem stems
  from       the       optimization       algorithm      (see       end       of
  subsection~\ref{SubSec:ThreeAxisPerformance}),  which  can converge  to  local
  extrema.  We can  circumvent this difficulty by  using optimization algorithms
  that  are  more  robust  against   local  extrema:  the  constrained  detector
  outperforms  the conventional  method  again,  but at  the  cost of  increased
  computational complexity,  which may  preclude real-time  application. Further
  research would be  necessary to develop a fast method  while limiting the risk
  of  getting stuck  at a  local extremum,  which is  beyond the  scope of  this
  article.
\end{itemize}

%\textcolor{red}{démo $\Pd{,E} \leq \Pd{,\V}$ ? Au secours Steeve !}

% We mentioned that signal coefficients exist in semi-algebraic spaces. However,
% certain  semi-algebraic  spaces can  be  viewed  as  ‘volumes’ as  opposed  to
% algebraic spaces, which can be viewed  as ‘surfaces’. This is particularly the
% case when  the coefficients  exist in  a full  elliptic half-cone.  Any signal
% whose coefficients  live in this  cone can physically  exist. But in  terms of
% detection, we believe that the closer  the coefficients are to the boundary of
% the cone,  the more  advantageous our  method becomes,  since once  noisy, the
% signal will be able to escape the  cone more easily and our method will remove
% even more noise.

% -------------------------------- Conclusion ---------------------------------- %

\section{Conclusion, discussion and perspectives}
\label{Sec:Conclusion}

We have examined various scenarios for  detecting magnetic anomalies. In each of
these scenarios, the signal evolves in a vector space of functions, which is not
always the smallest space containing all  possible signals. This behavior can be
explained  by the  fact  that  the basis  coefficients  depend  on the  physical
parameters that  constrain them, as  is the  case for vector  anomalies, squared
anomaly  modulus, and  the first  principal invariant  of the  magnetic gradient
tensor.   We have  also  shown that,  except  in very  specific  cases that  are
impractical,  a scalar  anomaly imposes  no constraints  on these  coefficients.
Therefore, from detection  purpose in the context of  the generalized likelihood
ratio  test,  it is  preferable  to  perform  the maximum  likelihood  estimator
directly on the  constrained space. The intuition behind this  is that it allows
for greater  noise filtering,  and robustness.  Through various  simulations, we
have  shown that  our  method  allows to  significantly  improve the  detector's
performance regardless of the configuration.

Still a  few issues  need to be  clarified or were  overlooked in  the present
study, and are deferred to a future paper.
\begin{itemize}
\item The  Gaussian additive noise  assumption is not always  strictly satisfied
  for all  examined detection  strategies. Actually, assuming  Gaussian additive
  noise  for case  of  vector measurement  of the  anomaly,  the square  modulus
  measurement under $\H_1$ reads:
  \begin{equation*}
  \|  \sb(\vartheta)  +   \nb  \|^2  =  \|  \sb(\vartheta)  \|^2   +  2  \langle
  \sb(\vartheta) | \nb \rangle + \| \nb \|^2
  \end{equation*}
  If  the  signal  was  whitened  beforehand,  then  $\|  \nb  \|^2$  follows  a
  chi-squared distribution and $\langle \sb(\vartheta)  | \nb \rangle$ follows a
  Gaussian distribution  dependent on $\vartheta$.  Although for a  large enough
  number of  degrees of  freedom, the Gaussian  approximation of  the Chi-square
  becomes  accurate,  this  needs  to  be  further  investigated  and  precisely
  analyzed.
\item Another challenge  involves estimating the angles $\alpha$  and $\beta$ in
  the vector  case; this leads  either to  the use of  computationally intensive
  methods to  obtain good  estimates, or to  less computationally  expensive but
  imperfect estimates  that may be stuck  in local extrema; this  is expected to
  degrade  the  receiver’s  performance.  An  optimization  procedure  combining
  robustness and speed  will therefore be necessary; for  example, an analytical
  approach—which remains to be developed—would likely be the most appropriate.
\item  Throughout the  article, the  source-CPA distance  was considered  known,
  which is  hardly the  case in  practice; The  statistical distribution  of its
  estimate will have to be incorporated into future studies of such detectors.
\end{itemize}

% -------------------------------- Appendix ---------------------------------- %

\appendix

% ------------------------ I1 along the trajector

\subsection{Second principal invariant along the trajectory}
\label{App:SecondInvariant}

If the second invariant is  considered for measurement, $\Pib(\Hb) = \det(\Gb)$;
Then      by     Eqs.~\eqref{Eq:uPrime}-\eqref{Eq:rTrajectory},      and     the
expression~\eqref{Eq:SecondInvariant}  ,  the  signature  $\sb$  accounting  for
sensor's trajectory reads:
\begin{equation*}
\sb(u) = \Ab \, \frac{1}{(1+u^2)^{\frac{15}{2}}} \: \begin{bmatrix}
1\\[1mm]
u\\[1mm]
u^2\\[1mm]
u^3
\end{bmatrix}
\end{equation*}
with
\begin{equation}
\label{Eq:CoeffSecondGradInv}
\Ab = - \: \frac{27}{D^{12}} \begin{bmatrix}
\big( {m_x'}^2 + {m_y'}^2 + 2 \, {m_z'}^2 \big) \, m_z'\\[3mm]
\big( {m_x'}^2 + {m_y'}^2 + 4 \, {m_z'}^2 \big) \, m_x' \\[3mm]
\big( 4 \, {m_x'}^2 + {m_y'}^2 + {m_z'}^2 \big) \, m_z' \\[3mm]
\big( 2 \, {m_x'}^2 + {m_y'}^2 + {m_z'}^2 \big) \, m_x'
\end{bmatrix}^\top
\end{equation}
The measured signals decomposes once again on a known basis $\F$, given by
\begin{equation*}
\label{Eq:BasisSecondGradInv}
\F = \left\{ u \mapsto \dfrac{u^i}{(1+u^2)^{\frac{15}2}} \right\}_{i=0}^3
\end{equation*}
with $\Ab  \equiv \Ab(\mb')$.  However, the  magnitude of  the signal  scales as
$\left( \frac{\|  \mb \|}{D^4} \right)^3$, making  the use of this  signal quite
difficult in practice.

%\Avoir{JE NE COMPRENDS  PAS The expression of invariant $2$  is provided so that
%  there are no  gaps in the racket, but  we will omit it from now  on because it
%  involves very complicated calculations.}

% ------------------------ Cone

\subsection{Identification of the space of physically feasible coefficients as a cone}
\label{App:HalfCone}

As   detailed  in~\cite{Sturm2017SurveyQuantifierElimination},   the  quantifier
elimination  algorithm does  not always  provide  a solution.   However, in  the
present case of interest, one can also proceed directly.

From~\eqref{Eq:AbSquareInvGeneric},  subtracting  the  first equality  from  the
third eliminates  $m_Y$: $(a-1) \,  {m_z'}^2 = (a-1) \,  {m_x'}^2 - A_z  + A_x$.
Then, squaring  the second equality and  using the previous result  we eliminate
$m_z'$, we obtain the second order equation in the variable ${m_x'}^2$:
\begin{equation*}
4 \, b^2 (a-1) \, {m_x'}^4 - 4 \, b^2 (A_z-A_x) \, {m_x'}^2 - (a-1) A_y^2 = 0
\end{equation*}
This equality has always a (positive) solution, given by
\begin{equation*}
{m_x'}^2 = \frac{b \, (A_z-A_x) + \sqrt{b^2 (A_z-A_x)^2 + (a-1)^2 A_y^2}}{2 \, b
  \, (a-1)}
\end{equation*}
and subsequently
\begin{equation*}
{m_z'}^2 = \frac{\sqrt{b^2 (A_z-A_x)^2 + (a-1)^2 A_y^2} - b \, (A_z-A_x)}{2 \, b
  \, (a-1)}
\end{equation*}
which  always exists.  Finally, both  the first
and the last equation in Eq.~\eqref{Eq:AbSquareInvGeneric} gives
\begin{equation*}
{m_y'}^2  =  \frac{A_z+A_x}{2} -  \frac{(a+1)  \sqrt{b^2  (A_z-A_x)^2 +  (a-1)^2
    A_y^2}}{2 \, b \, (a-1)}
% \frac{ b (a-1) (A_z+A_x) - (a+1)  \sqrt{b^2 (A_z-A_x)^2 + (a-1)^2 A_y^2}}{2 \,
% b \, (a-1)}
\end{equation*}
which must  be positive.  This requirement  leads to the cone  equation together
with $A_x + A_z \geqslant 0$.  The latter inequality restricts to $A_z \geqslant
0$ since,  before the rotation  about the $y-$axis of  the cone is  applied, the
semi $x-$axis of the  elliptical base is $$ \frac{a-1}{a+1} z  < z$$ thus, after
the rotation, $A_x$ has the same sign as $A_z$.

% ------------------------ Quartic polynomial
  
\subsection{Quartic polynomial}
\label{App:QuarticPolynomial}

By   substituting   Eq.~\eqref{Eq:HA_Vperp}   in    second   equation   of   the
system~\eqref{Eq:LagrangianSystem},  $\lambda_\V$ is  solution of  the following
equation:
\begin{equation}
\label{Eq:LagrangeParameter}
\overline{\Ab}_\F  \left( \Ib_3  + \lambda_\V  \overline{\Qb}_{a,b} \right)^{-1}
\overline{\Qb}_{a,b} \left( \Ib_3 + \lambda_\V \overline{\Qb}_{a,b} \right)^{-1}
\overline{\Ab}_\F = 0
\end{equation}
with\footnote{$\Mb^{\frac12}$ is the unique  non-negative definite matrix square
root  of the  non-negative definite  matrix $\Mb$~\cite{Meyer2200MatrixAnalysis,
  Gantmacher1959TheorieMatricesV1}.}
\begin{equation}\left\{\begin{array}{lll}
\overline{\Ab}_\F & = & \Ab_\F \left( \tFb \, \tFb^\top\right)^{\frac12}\\[2mm]
\overline{\Qb}_{a,b} & = &  \left( \tFb \, \tFb^\top\right)^{-\frac12} \Qb_{a,b}
\left( \tFb \, \tFb^\top\right)^{-\frac12}
\end{array}\right.\end{equation}
        
One observes then  that the characteristic polynomials of a  $3 \times 3$ matrix
$\Qb$
\begin{equation*}
P_{\Qb}(x) =  \det\left( \Qb - x  \, \Ib_3 \right) =  - x^3 + I_0(\Qb)  \, x^2 +
I_1(\Qb) \, x + I_2(\Qb)
\end{equation*}
where
\begin{equation*}\left\{\begin{array}{lll}
I_0(\Qb) & = & \displaystyle \Tr(\Qb)\\[2mm]
I_1(\Qb)  & =  &  \displaystyle \frac12  \left(  \Tr\left(\Qb^2\right) -  \left(
\Tr(\Qb) \right)^2 \right)\\[3mm]
I_2(\Qb) &  = & \displaystyle \det(\Qb)
  \end{array}\right.\end{equation*}
are            the             principal            invariants            evoked
previously~\cite{Gantmacher1959TheorieMatricesV1,  Hou1998ProofLeverrierFaddeev,
  Spencer2004ContinuumMechanics}.      Now,     by     using     Cayley-Hamilton
theorem~\cite{Meyer2200MatrixAnalysis,   Gantmacher1959TheorieMatricesV1},   the
matrix   zeroes   its   characteristic   polynomials,   thus   when   $\Qb$   is
invertible,
\begin{equation}
\label{Eq:CayleyHamilton}
\det(\Qb) \, \Qb^{-1} = \Qb^2 - I_0(\Qb) \, \Qb - I_1(\Qb) \, \Ib_3
\end{equation}
Injecting  now   $\Qb  =   \Ib_3  +   \lambda_\V  \,   \overline{\Qb}_{a,b}$  in
Eqs~\eqref{Eq:CayleyHamilton}-\eqref{Eq:LagrangeParameter}, the result follows.

\bibliographystyle{unsrt}%plain}
\bibliography{bibliography.bib}

@book{Abramowitz1972handbook,
  author    = {Abramowitz, Milton and Stegun, Irene A.},
  title     = {Handbook of mathematical functions with formulas, graphs, and mathematical tables},
  edition   = {10th},
  year      = {1972},
  address   = {New-York, USA},
  publisher = {Dover Publication}
}

@inproceedings{Basu1994ComplexityQuantifier,
  author    = {Basu, Saugata and  Pollack, Richard and Roy, M.-F.},
  title     = {On the combinatorial and algebraic complexity of quantifier elimination},
  booktitle = {Proceedings 35th Annual Symposium on Foundations of Computer Science},
  pages     = {632--641},
  year      = {1994},
  publisher = {IEEE}
}

@phdthesis{Blanpain1979PhD,
  author = {Blanpain, Roland},
  title  = {Traitement en temps réel du signal issu d’une sonde magnétométrique pour la détection d’anomalies magnétiques},
  year   = {1979},
  school = {INP Grenoble},
  type   = {PhD thesis}
}

@book{Boyd2004ConvexOptimization,
  author    = {Boyd, Stephen and Vandenberghe, Lieven},
  title     = {Convex Optimization},
  year      = {2004},
  address   = {Cambridge, UK},
  publisher = {Cambridge University Press}
}

@article{Byrd1995limited,
  author    = {Byrd, Richard H. and Lu, Peihuang and Nocedal, Jorge and Zhu, Ciyou},
  title     = {A limited memory algorithm for bound constrained optimization},
  journal   = {SIAM Journal on scientific computing},
  volume    = {16},
  number    = {5},
  pages     = {1190--1208},
  year      = {1995},
  publisher = {SIAM}
}

@article{Chen2024magnetic,
  author    = {Chen, Zhikun and Lou, Yuchao and He, Pengfei and Xu, Pengcheng and Zhang, Xiaofeng},
  title     = {Magnetic Anomaly Detection Based on Attention-Bi-{LSTM} Network},
  journal   = {IEEE Transactions on Instrumentation and Measurement},
  volume    = {73},
  pages     = {1--11},
  year      = {2024},
  publisher = {IEEE}
}

@inproceedings{Chenevasdetection,
  author = {Chenevas-Paule, Cl{\'e}ment and Zozor, Steeve and Rouve, Laure-Line and Michel, Olivier and Pinaud, Olivier and Kukla, Romain},
  title  = {D{\'e}tection d’anomalies magn{\'e}tiques sous contraintes physiques},
  booktitle    = {30th Colloque Colloque Francophone de Traitement du Signal et des Images (GRETSI)},
  year         = {2025},
  organization = {GRETSI}
}

@inproceedings{Chenevas2024analytical,
  author       = {Chenevas-Paule, Cl{\'e}ment and Zozor, Steeve and Rouve, L.-L. and Michel, Olivier J.-J. and Pinaud, Olivier and Kukla, Romain},
  title        = {On an analytical orthonormal multipolar basis for magnetic anomaly detection},
  booktitle    = {32nd European Signal Processing Conference (EUSIPCO)},
  pages        = {2567--2571},
  year         = {2024},
  organization = {IEEE}
}

@article{Chenevas2026MAD,
  author    = {Chenevas-Paule, Cl{\'e}ment and Zozor, Steeve and Rouve, Laure-Line and Michel, Olivier JJ and Pinaud, Olivier and Kukla, Romain},
  title     = {On multipolar magnetic anomaly detection: multipolar signal subspaces, an analytical orthonormal basis, multipolar truncation and detection performance},
  journal   = {EURASIP Journal on Advances in Signal Processing (to appear)},
  year      = {2026},
  publisher = {Springer}
}

@inproceedings{Chenevas2025physics,
  author       = {Chenevas-Paule, Cl{\'e}ment and Zozor, Steeve and Rouve, Laure-Line and Michel, Olivier J.-J. and Pinaud, Olivier and Kukla, Romain},
  title        = {On physics-constrained magnetic anomaly detection},
  booktitle    = {23rd IEEE Statistical Signal Processing Workshop (SSP)},
  pages        = {291--295},
  year         = {2025},
  organization = {IEEE}
}

@article{Colson2007Overviewbilevel,
  author    = {B. Colson and P. Marcotte and G. Savard},
  title     = {An overview of bilevel optimization},
  journal   = {Annals of Operations Research},
  volume    = {153},
  number    = {1},
  pages     = {235--256},
  year      = {2007},
  publisher = {Springer}
}

@book{Coste2000introduction,
  author    = {Coste, Michel},
  title     = {An introduction to semialgebraic geometry},
  year      = {2000},
  address   = {Pisa, Italy},
  publisher = {Istituti editoriali e poligrafici internazionali}
}

@article{Davenport1988QuantifierExponential,
  author  = {Davenport, James H. and Heintz, Joos},
  title   = {Real quantifier elimination is doubly exponential},
  journal = {Journal of Symbolic Computation},
  volume  = {5},
  number  = {1–2},
  pages   = {29--35},
  year    = {1988},
  publisher = {Elsevier}
}

@book{Dempe2002Bilevelprogramming,
  author    = {Dempe, Stephan},
  title     = {Foundations of Bilevel Programming},
  year      = {2002},
  address   = {Dordrecht, The Netherlands},
  publisher = {Kluwer}
}

@article{Dolzmann1997redlog,
  author    = {Dolzmann, Andreas and Sturm, Thomas},
  title     = {Redlog: Computer algebra meets computer logic},
  journal   = {ACM SIGSAM Bulletin},
  volume    = {31},
  number    = {2},
  pages     = {2--9},
  year      = {1997},
  publisher = {ACM}
}

@article{Fan2020adaptive,
  author    = {Fan, Liming and Kang, Chong and Wang, Huigang and Hu, Hao and Zhang, Xiaojun and Liu, Xing},
  title     = {Adaptive magnetic anomaly detection method using support vector machine},
  journal   = {IEEE Geoscience and Remote Sensing Letters},
  volume    = {19},
  pages     = {1--5},
  year      = {2020},
  publisher = {IEEE}
}

@article{Fan2020gradient,
  author    = {Fan, Liming and Kang, Chong and Hu, Hao and Zhang, Xiaojun and Liu, Jianguo and Liu, Xing and Wang, Huigang},
  title     = {Gradient signals analysis of scalar magnetic anomaly using orthonormal basis functions},
  journal   = {Measurement Science and Technology},
  volume    = {31},
  number    = {11},
  pages     = {115105},
  year      = {2020},
  publisher = {IOP Publishing}
}

@proceedings{Fitterman1987MADSurveys,
  editor    = {Fitterman, David V.},
  title     = {Proceedings of the {U. S.} Geological Survey Workshop on the Development and Application of Modern Airborne Electromagnetic Surveys},
  year      = {1987},
  address   = {Denver, CO, USA},
  publisher = {{U. S.} Geological Survey Bulletin 1925}
}

@book{Gantmacher1959TheorieMatricesV1,
  author    = {Gantmacher, Felix R.},
  title     = {The Theory of Matrices - Volume 1},
  year      = {1959},
  address   = {New-York, USA},
  publisher = {Chelsea Publishing Company}
}

@article{Ginzburg2002processing,
  author    = {Ginzburg, Boris and Frumkis, Lev and Kaplan, Ben-Zion},
  title     = {Processing of magnetic scalar gradiometer signals using orthonormalized functions},
  journal   = {Sensors and Actuators A: Physical},
  volume    = {102},
  number    = {1-2},
  pages     = {67--75},
  year      = {2002},
  publisher = {Elsevier}
}

@book{Gupta2018matrix,
  author    = {Gupta, Arjun K. and Nagar, Daya K.},
  title     = {Matrix variate distributions},
  year      = {2018},
  address   = {Boca Rato, FL, USA},
  publisher = {Chapman and Hall/CRC}
}

@PhdThesis{Hezel2020ImprovingCalibration,
  author  = {Hezel, Mitchel},
  title   = {Improving Aeromagnetic Calibration Using Neural Networks},
  school  = {Air Force Institute of Technology, Wright-Patterson Air Force Base},
  year    = {2020},
  address = {Wright-Patterson Air Force Base, Ohio, USA},
}

@article{Hou1998ProofLeverrierFaddeev,
  author  = {Hou, Shiu-Hung},
  title   = {A Simple Proof of the {L}everrier--{F}addeev Characteristic Polynomial Algorithm},
  journal = {SIAM Review},
  volume  = {40},
  number  = {3},
  pages   = {706--709},
  year    = {1998}
}

@article{Hu2020magnetic,
  author    = {Hu, Mengkai and Jing, Sen and Du, Changping and Xia, Mingyao and Peng, Xiang and Guo, Hong},
  title     = {Magnetic dipole target signal detection via convolutional neural network},
  journal   = {IEEE Geoscience and Remote Sensing Letters},
  volume    = {19},
  pages     = {1--5},
  year      = {2020},
  publisher = {IEEE}
}

@book{Johnson1995:v1,
  author    = {Johnson, Norman L. and Kotz, Samuel and Balakrishnan, Narayanaswamy},
  title     = {Continuous Univariate Distributions},
  volume    = {1},
  edition   = {2nd},
  year      = {1995},
  address   = {New-York, USA},
  publisher = {John Wiley {\&} Sons}
}

@book{Kay1998detection,
  author    = {Kay, Stephen M.},
  title     = {Fundamentals for Statistical Signal Processing: Detection Theory},
  volume    = {2},
  year      = {1998},
  address   = {Englewood Cliffs, NJ, USA},
  publisher = {Prentice Hall}
}

@article{Koch1984Matrixinvariants,
  author  = {Koch, Richard},
  title   = {Matrix Invariants},
  journal = {The American Mathematical Monthly},
  volume  = {91},
  number  = {9},
  pages   = {573--575},
  year    = {1984},
  organization = {Taylor \& Francis}
}

@book{Lasserre2015PolynomialOptimization,
  author    = {Lasserre, Jean Brenard},
  title     = {Introduction to Polynomial and Semi-Algebraic Optimization},
  year      = {2015},
  address   = {Cambridge, UK},
  publisher = {Cambridge University Press}
}

@article{Lelial1961Identification,
  author    = {Leliak, Paul},
  journal   = {IRE Transactions on Aeronautical and Navigational Electronics},
  title     = {Identification and Evaluation of Magnetic-Field Sources of Magnetic Airborne Detector Equipped Aircraft},
  volume    = {ANE-8},
  number    = {3},
  pages     = {95--105},
  year      = {1961},
  publisher = {IEEE}
}

@article{Liu2019magnetic,
  author    = {Liu, Shuchang and Chen, Zhuo and Pan, Mengchun and Zhang, Qi and Liu, Zhongyan and Wang, Siwei and Chen, Dixiang and Hu, Jingtao and Pan, Xue and Hu, Jiafei and Li, Peisen and Wan, Chengbiao},
  title     = {Magnetic anomaly detection based on full connected neural network},
  journal   = {IEEE Access},
  volume    = {7},
  pages     = {182198--182206},
  year      = {2019},
  publisher = {IEEE}
}

@techreport{Loane1976speed,
  author      = {Loane, Edward P.},
  title       = {Speed and depth effects in magnetic anomaly detection},
  institution = {Defense Technical Information Center},
  number      = {ADA081329, unclassified},
  year        = {1976},
  address     = {EPL Analysis, Ashton, MD, USA}
}

@book{Magnus1999MatrixDifferential,
  author    = {Magnus, Jan R. and Neudecker, Heinz},
  title     = {Matrix Differential Calculus with Applications in Statistics and Econometrics},
  edition   = {3rd},
  year      = {1999},
  address   = {New-York},
  publisher = {John Wiley \& Sons}
}

@book{Meyer2200MatrixAnalysis,
  author    = {Meyer, Carl D.},
  title     = {Matrix Analysis and Applied Linear Algebra},
  year      = {2000},
  address   = {Philadelphia, PA, USA},
  publisher = {SIAM}
}

@inproceedings{Nerrise2023physics,
  author    = {Nerrise, Favour and Sosanya, Andrew Sosa and Neary, Patrick},
  title     = {Physics-informed calibration of aeromagnetic compensation in magnetic navigation systems using liquid time-constant networks},
  booktitle = {37th Conference on Neural Information Processing (NeurIPS)},
  month     = {10-16 december},
  year      = {2023},
  address   = {New Orleans, LA, USA}
}

@article{Patnaik1949NonCentral,
  author    = {Patnaik, P. B.},
  title     = {The Non-Central {$\chi^2$}- and {$F$}-Distribution and their Applications},
  journal   = {Biometrika},
  volume    = {36},
  number    = {1/2},
  pages     = {202--232},
  year      = {1949},
  publisher = {Oxford Academic}
}

@inproceedings{Pepe2015generalization,
  author    = {Pepe, Pascal and Zozor, Steeve and Rouve, Laure-Line and Coulomb, Jean-Louis and Servi{\`e}re, Christine and Muley, Jean},
  title     = {Generalization of GLRT-based magnetic anomaly detection},
  booktitle = {23rd European Signal Processing Conference (EUSIPCO)},
  pages     = {1930--1934},
  year      = {2015},
  publisher = {IEEE}
  }

@article{Qiao2023adaptive,
  author    = {Qiao, Shuai and Wang, Qimeng and Zheng, Doudou and Hou, Qingfeng and Zhao, Junzhi and Tang, Jun and Yanjun, Li and Sugawara, Yasuhiro and Ma, Zongmin and Liu, Jun},
  title     = {Adaptive filter entropy monitoring method for scalar magnetic detection using optically pumped magnetometers},
  journal   = {Measurement Science and Technology},
  volume    = {34},
  number    = {5},
  pages     = {055107},
  year      = {2023},
  publisher = {IOP Publishing}
}

@article{Qin2020magnetic,
  author    = {Qin, Yijie and Li, Keyan and Yao, Chang and Wang, Xianran and Ouyang, Jun and Yang, Xiaofei},
  title     = {Magnetic anomaly detection using full magnetic gradient orthonormal basis function},
  journal   = {IEEE Sensors Journal},
  volume    = {20},
  number    = {21},
  pages     = {12928--12940},
  year      = {2020},
  publisher = {IEEE}
}

@article{Sheinker2009magnetic,
  author    = {Sheinker, Arie and Frumkis, Lev and Ginzburg, Boris and Salomonski, Nizan and Kaplan, Ben-Zion},
  title     = {Magnetic anomaly detection using a three-axis magnetometer},
  journal   = {IEEE Transactions on Magnetics},
  volume    = {45},
  number    = {1},
  pages     = {160--167},
  year      = {2009},
  publisher = {IEEE}
}

@article{Sheinker2008magnetic,
  author    = {Sheinker, Arie and Salomonski, Nizan and Ginzburg, Boris and Frumkis, Lev and Kaplan, Ben-Zion},
  title     = {Magnetic anomaly detection using entropy filter},
  journal   = {Measurement science and technology},
  volume    = {19},
  number    = {4},
  pages     = {045205},
  year      = {2008},
  publisher = {IOP Publishing}
}

@book{Spencer2004ContinuumMechanics,
  author    = {Spencer, Anthony J. M.},
  title     = {Continuum Mechanics},
  year      = {2004},
  address   = {Mineola, New-York, USA},
  publisher = {Dover Publications}
}

@book{Straton2007Electromagnetic,
  author    = {Stratton, Julius Adams},
  title     = {Electromagnetic Theory},
  year      = {2007},
  address   = {Hoboken, NJ, USA},
  publisher = {Wiley-IEEE Press}
}

@article{Sturm2017SurveyQuantifierElimination,
  author    = {Sturm, Thomas},
  title     = {A Survey of Some Methods for Real Quantifier Elimination, Decision, and Satisfiability and Their Applications},
  journal   = {Mathematics in Computer Science},
  volume    = {11},
  number    = {3–4},
  pages     = {483--502},
  year      = {2017},
  publisher = {Springer}
}

@book{vanTrees2013DetectionEstimation,
  author    = {{Van Trees}, Harry L. and Bell, Kristine L. and Thian, Zhi},
  publisher = {John Wiley {\&} Sons},
  title     = {Detection Estimation and Modulation Theory. Part I: Detection, Estimation, and Filtering Theory},
  year      = {2013},
  address   = {Hoboken, New Jersey, USA},
  edition   = {2nd}
}

@article{Wang2022deep,
  author    = {Wang, Yizhen and Han, Qi and Zhao, Guanyi and Li, Minghui and Zhan, Dechen and Li, Qiong},
  title     = {A deep neural network based method for magnetic anomaly detection},
  journal   = {IET Science, Measurement {\&} Technology},
  volume    = {16},
  number    = {1},
  pages     = {50--58},
  year      = {2022},
  publisher = {IET}
}

@article{Wikwo1984Multipole,
  author  = {Wikswo, John P. and Swinney, Kenneth R.},
  title   = {A comparison of scalar multipole expansions},
  journal = {Journal of Applied Physics},
  volume  = {56},
  number  = {11},
  pages   = {3039-3049},
  year    = {1984},
  publisher = {AIP Publishing}
}

@incollection{Wynn1999DetectionStaticDipole,
  author    = {Wynn, W. Michael},
  title     = {Detection, Localization and Characterization of Static Magnetic-Dipole Source},
  booktitle = {Detection And Identification Of Visually Obscured Targets},
  chapter   = {11},
  editor    = {Baum, Carl E.},
  year      = {1999},
  address   = {Boca Raton, FL, USA},
  publisher = {CRC Press},
}

@article{Xu2020deepmad,
  author    = {Xu, Xin and Huang, Ling and Liu, Xiaojun and Fang, Guangyou},
  journal   = {IEEE Access},
  title     = {Deep{MAD}: Deep learning for magnetic anomaly detection and denoising},
  volume    = {8},
  pages     = {121257--121266},
  year      = {2020},
  publisher = {IEEE}
}

@article{Yan2024effective,
  author    = {Yan, Youyu and Liu, Jianguo and Yan, Shenggang and Shen, Siyuan and Li, Xiangang},
  title     = {An effective magnetic anomaly detection using orthonormal basis of magnetic gradient tensor invariants},
  journal   = {IEEE Transactions on Geoscience and Remote Sensing},
  volume    = {62},
  pages     = {1--11},
  year      = {2024},
  publisher = {IEEE}
}

@article{Yang2018magnetic,
  author    = {Yang, Liu and Zhongyan, Liu and Mengchun, Pan and Qi, Zhang and Dixiang, Chen and Chengbiao, Wan and Gui, Hu and Dewen, Zhang and Zhuo, Chen},
  title     = {Magnetic anomaly signal space analysis and its application in noise suppression},
  journal   = {IEEE Geoscience and Remote Sensing Letters},
  volume    = {16},
  number    = {1},
  pages     = {130--134},
  year      = {2018},
  publisher = {IEEE}
}

@article{Zhao2021MADReview,
  author  = {Zhao, Yue and Zhang, Junhai and Li, Jiahui and Lui, Shuangqiang and Miao, Peixian and Shi, Yanchao and Zhao, Enming}, 
  title   = {A brief review of magnetic anomaly detection},
  journal = {Measurement Science and Technology},
  volume  = {32},
  number  = {4},
  pages   = {042002},
  year    = {2021},
  publisher={IOP Publisher}
}

@article{Zhu1997algorithm,
  title     = {Algorithm 778: {L-BFGS-B}: Fortran subroutines for large-scale bound-constrained optimization},
  author    = {Zhu, Ciyou and Byrd, Richard H and Lu, Peihuang and Nocedal, Jorge},
  journal   = {ACM Transactions on mathematical software},
  volume    = {23},
  number    = {4},
  pages     = {550--560},
  year      = {1997},
  publisher = {ACM}
}

@techreport{MADProgram1946,
  address     = {Washington, D.C. USA},
  institution = {Summary technical report of the National Defense Research Committee, Division 6},
  title       = {Magnetic Airborne Detection Program},
  volume      = {5},
  number      = {AD221590, unclassified},
  year        = {1946}
}

\end{document}